\pdfoutput=1
\documentclass[preprint,prd,aps,preprintnumbers,nofootinbib,superscriptaddress,showpacs,floatfix]{revtex4}
\usepackage{amsmath,amssymb,amsbsy}
\usepackage{graphicx}
\usepackage{epsfig}
\usepackage{color}

\newcommand{\Delstar}{\ensuremath{\Delta^{\raise0.18ex\hbox{${\scriptstyle *}$}}}}
\def\gtwid{{\,\raise.35ex\hbox{$>$\kern-.75em\lower1ex\hbox{$\sim$}}\,}}
\def\ltwid{{\,\raise.35ex\hbox{$<$\kern-.75em\lower1ex\hbox{$\sim$}}\,}}
\def\leftvec{{\raise1.5ex\hbox{$\leftarrow$}\kern-1.00em}}
\def\rightvec{{\raise1.5ex\hbox{$\rightarrow$}\kern-1.00em}}
\def\half{{\scriptstyle \raise.2ex\hbox{${1\over2}$}}}
\def\threehalves{{\scriptstyle \raise.15ex\hbox{${3\over2}$}}}
\def\third{{\scriptstyle \raise.15ex\hbox{${1\over3}$}}}
\def\third{{\scriptstyle \raise.15ex\hbox{${1\over3}$}}}
\def\twothirds{{\scriptstyle \raise.15ex\hbox{${2\over3}$}}}
\def\fourth{{\scriptstyle \raise.15ex\hbox{${1\over4}$}}}

\newcommand*{\bea}{\begin{eqnarray}}
\newcommand*{\eea}{\end{eqnarray}}
\newcommand*{\be}{\begin{equation}}
\newcommand*{\ee}{\end{equation}}

\newcommand*{\CPT}{\raise0.4ex\hbox{$\chi$}PT}
\newcommand*{\chpt}{\raise0.4ex\hbox{$\chi$}PT}
\newcommand*{\schpt}{S\raise0.4ex\hbox{$\chi$}PT}

\def\eqref#1{{(\ref{#1})}}

\def\bar{\overline}
\def\hat{\widehat}

\def\bea{\begin{eqnarray}}
\def\eea{\end{eqnarray}}

\def\beq{\begin{equation}}
\def\eeq{\end{equation}}

\def\spose#1{\hbox to 0pt{#1\hss}}
\def\ltapprox{\mathrel{\spose{\lower 3pt\hbox{$\mathchar"218$}}
 \raise 2.0pt\hbox{$\mathchar"13C$}}}
\def\gtapprox{\mathrel{\spose{\lower 3pt\hbox{$\mathchar"218$}}
 \raise 2.0pt\hbox{$\mathchar"13E$}}}
\def\inapprox{\mathrel{\spose{\lower 3pt\hbox{$\mathchar"218$}}
 \raise 2.0pt\hbox{$\mathchar"232$}}}

\begin{document}

\preprint{IUHET-535}
\preprint{0910.2928}

\vphantom{}

\title{Lattice QCD inputs to the CKM unitarity triangle analysis}

\author{Jack Laiho}
\email[]{jlaiho@fnal.gov}
\affiliation{Physics Department, Washington University, St. Louis, MO 63130}
\affiliation{Department of Physics and Astronomy, University of Glasgow, Glasgow, G128 QQ, UK}

\author{E.\ Lunghi}
\email[]{elunghi@indiana.edu}
\affiliation{Physics Department, Indiana University, Bloomington, IN 47405}

\author{Ruth S. Van de Water}
\email[]{ruthv@bnl.gov}
\affiliation{Physics Department, Brookhaven National Laboratory, Upton, NY 11973}

%\date{\today}

%=================================================
%The abstract
%=================================================
\begin{abstract}
We perform a global fit to the CKM unitarity triangle using the latest experimental and theoretical constraints.  Our emphasis is on the hadronic weak matrix elements that enter the analysis, which must be computed using lattice QCD or other nonperturbative methods.  Realistic lattice QCD calculations which include the effects of the dynamical up, down, and strange quarks are now available for all of the standard inputs to the global fit.  We therefore present lattice averages for all of the necessary hadronic weak matrix elements.  We attempt to account for correlations between lattice QCD results in a reasonable but conservative manner:  whenever there are reasons to believe that an error is correlated between two lattice calculations, we take the degree of correlation to be 100\%.  These averages are suitable for use as inputs both in the global CKM unitarity triangle fit and other phenomenological analyses.  In order to illustrate the impact of the lattice averages, we make Standard Model predictions for the parameters $\hat{B}_K$, $|V_{cb}|$, and $|V_{ub}|/|V_{cb}|$.  We find a (2-3)$\sigma$ tension in the unitarity triangle, depending upon whether we use the inclusive or exclusive determination of $|V_{cb}|$.  If we interpret the tension as a sign of new physics in either neutral kaon or $B$-mixing, we find that the scenario with new physics in kaon-mixing is preferred by present data.
\end{abstract}

\maketitle

\section{Introduction}

The B-factories and the Tevatron have been remarkably successful in producing a wealth of data needed to constrain the flavor sector of the Standard Model.  Inconsistencies between independent determinations of the elements of the Cabibbo-Kobayashi-Maskawa (CKM) matrix~\cite{Cabibbo:1963yz,Kobayashi:1973fv} and its CP-violating phase would provide evidence for new physics.  The search for such inconsistencies has received a great deal of attention because generic new physics scenarios lead to additional CP-violating phases beyond the single one of the Standard Model.  Physicists generally believe that the Standard Model cannot be the whole story, despite its great experimental successes, because it does not describe the large matter-antimatter asymmetry of the universe, nor can it account for dark matter.  Although there is reasonably good agreement with the Standard Model prediction of a single CP-violating phase, as encoded in the CKM matrix, some tensions have been pointed out recently~\cite{Lunghi:2007ak,Lunghi:2008aa,Buras:2008nn,Buras:2009pj,Lunghi:2009sm}.  

Most of the constraints on the CKM matrix are limited by theoretical uncertainties in the hadronic matrix elements that encode the nonperturbative QCD contributions to weak processes.  These hadronic matrix elements are calculated using numerical lattice QCD, and great progress has been made in reducing the errors in the last few years.\footnote{\textcolor{black}{For a pedagogical review of lattice QCD methods see Ref.~\cite{Bazavov:2009bb}}.} Three flavors of light quarks ($u$, $d$, and $s$) are now being included in the vacuum polarization, so that the quenched approximation has become a thing of the past for most quantities of interest.  Lattice calculations of many quantities have now been done by various groups with all sources of systematic error under control.\footnote{\textcolor{black}{See, for example, Refs.~\cite{Gamiz:2008bd,Lellouch:2009fg}, for the status of lattice calculations of kaon and heavy-light physics.}}  In order to maximize the impact of lattice input on phenomenology, it is necessary to average the different results.  This is not entirely straightforward, since correlations between various lattice errors must be taken into account.  For example, statistical errors in quantities computed on the same gauge ensembles will be highly correlated.  Systematic errors can also be correlated, so familiarity with the lattice methods used in each calculation is needed in order to understand and account for these correlations in the averaging procedure.  

In this work we present lattice QCD averages of the hadronic weak matrix elements that enter the global fit of the CKM unitarity triangle.  We provide results for the neutral kaon mixing parameter ($B_K$), for the neutral $B$-meson decay constants and mixing matrix elements, for the inclusive determinations of the CKM matrix elements $|V_{ub}|$ and $|V_{cb}|$, for the Standard Model correction to $\epsilon_K$ ($\kappa_\epsilon$), and for the kaon decay constant ($f_K$).  Although we do not know the exact correlations between different lattice calculations of the same quantity, we still attempt to account for correlations in a reasonable manner.  We do so by assuming that, whenever a source of error is at all correlated between two lattice calculations, the degree-of-correlation is 100\%.  This assumption is conservative and will lead to somewhat of an overestimate in the total error of the lattice averages;  nevertheless, it is the most systematic treatment possible without knowledge of the correlation matrices between the various calculations, which do not exist.  These averages are intended for use in the global CKM unitarity triangle fit, as well as other phenomenological analyses.\footnote{\textcolor{black}{There are some additional correlations between different lattice quantities that enter the unitarity triangle fit, since collaborations often calculate more than one quantity using the same gauge configurations.  Although we do not include these effects in our analysis, such correlations are reduced in the procedure of averaging several lattice results for the same quantity that have been computed with different configurations.  Therefore these correlations should become less significant over time as more independent $N_f = 2+1$ flavor lattice results become available to include in the averages.}} 

Our emphasis in this work is different than that of the CKMfitter and UTfit collaborations~\cite{CKMfitter,UTFit}.  Their focus is primarily on the statistical techniques used to extract information from a given set of inputs (\textcolor{black}{see Ref.~\cite{Battaglia:2003in} for a quantitative comparison of the Bayesian versus frequentist approaches}), whereas our focus is on the lattice QCD inputs themselves.  Nevertheless, it is useful to point out some important differences between the manner in which they compute their lattice averages and the treatment that we use in this paper.  Both CKMfitter and UTfit combine two- and three-flavor results in their lattice averages~\cite{CKMfitter_lat,UTFit_lat}, which we do not.  This is because, although the systematic uncertainty in a particular quantity due to omitting the dynamical strange quark may in fact be small, this is impossible to quantify until the equivalent calculation has been done with both two and three flavors.  We therefore only consider $N_f = 2+1$ lattice results in our averages.  CKMfitter assigns the smallest systematic error of any of the individual lattice calculations to the average~\cite{CKMfitter_lat}  instead of combining the systematic uncertainties between different lattice calculations.  This treatment does not take full advantage of current lattice QCD results by preventing the average value from having a smaller systematic uncertainty than any of the individual values.  Thus, although this treatment is conservative, it may obscure the presence of new physics in an attempt to be overly cautious.  The method for obtaining the central values and errors used by UTfit is not fully spelled out in Ref.~\cite{UTFit_lat}.   In this work we take all quoted lattice errors at face value when averaging results, but we only include results with complete systematic error budgets that have been sufficiently documented in either a publication or conference proceeding.

This paper is organized as follows.  In Sec.~\ref{sec:LatticeInputs} we average lattice QCD results for hadronic weak matrix elements that enter the standard global fit of the CKM unitarity triangle.  We present the individual results used in the averages, briefly describe the methods used in each lattice calculation, and spell out which errors we consider correlated between the calculations. Next, for completeness, we summarize the other inputs used in our unitarity triangle fit in Sec.~\ref{sec:OtherInputs}.  We then illustrate the impact of the new lattice averages on the unitarity triangle fit in Sec.~\ref{sec:SMpredictions}.  We observe a (2--3)$\sigma$ tension in the fit, depending upon whether we use the inclusive or exclusive determination of $|V_{cb}|$.  In Sec.~\ref{sec:NewPhys} we interpret the tension as a sign of new physics in either neutral kaon or $B$-meson mixing.  We find that the current data prefer the scenario in which the new physics is in the kaon sector.  Finally, we summarize our results and conclude in Sec.~\ref{sec:Conc}.  \textcolor{black}{Summary plots of all of the lattice averages are provided in Appendix~\ref{sec:App}.}

%=================================================
\section{Lattice QCD inputs to the fit of the unitarity triangle}
\label{sec:LatticeInputs}
%=================================================

Many of the constraints on the CKM unitarity triangle rely upon knowledge of hadronic matrix elements that parameterize the nonperturbative QCD contributions to weak decays and mixing.  In the past, these hadronic weak matrix elements have often been difficult to compute precisely, and have enabled only mild ($10-15\%$ level) constraints on the apex of the CKM unitarity triangle that have been insufficient to probe the presence of new physics.  Recent advances in computers, algorithms, and actions, however, now allow reliable lattice QCD calculations of hadronic weak matrix elements with all sources of systematic error under control.  State-of-the-art lattice computations now regularly include the effects of the dynamical up, down, and strange quarks.  They also typically simulate at pion masses below 300 MeV, and sometimes even below 200 MeV, in order to control the extrapolation to the physical pion mass.  For many hadronic weak matrix elements of interest, there are now at least two reliable lattice calculations.  Just as with experimental measurements, some of the errors are correlated among the lattice QCD calculations, and such correlations much be taken into account when averaging lattice inputs to be used in the CKM unitarity triangle analysis.

In this section, we average the latest lattice QCD results and provide values that should be used in current fits of the CKM unitarity triangle.  In the averages, we only include results from simulations with three dynamical quark flavors, and with associated proceedings or publications that include comprehensive error budgets.  Fortunately, for all quantities of interest, there is at least one calculation that satisfies these critera.  In taking the averages we assume that all errors are normally distributed and follow the prescription outlined in Ref.~\cite{Schmelling:1994pz} to take the correlations into account. The degree of correlation induced by a given source of uncertainty onto the errors of different lattice calculations is extremely difficult to estimate. In order to be conservative, whenever there are arguments that suggest some correlation between errors in distinct lattice results, we take it to be 100\%. Finally, we adopt the PDG prescription to combine several measurements whose spread is wider than what expected from the quoted errors: the error on the average is increased by the square root of the minimum of the chi-square per degree of freedom (constructed following Ref.~\cite{Schmelling:1994pz}). 

%=================================================
\subsection{$B_K$}
\label{sec:BK}
%=================================================

The experimental measurement of indirect CP-violation in the kaon sector, $\varepsilon_K$, when combined with a nonperturbative determination of the neutral kaon mixing parameter, $B_K$, places a constraint on the apex of the unitarity triangle.  There have been three realistic lattice QCD calculations of $B_K$ since 2006;  the results are summarized in Table~\ref{tab:LQCD_BK}.

\begin{table}[t]
\begin{center}
\begin{tabular}{lccc}
 \hline \hline
 & $\hat B_K$ & $\;\;\;\;( \delta \hat B_K)_{\rm stat} $ &  $\;\;\;\;( \delta \hat B_K)_{\rm syst}$ \\
HPQCD/UKQCD '06~\cite{Gamiz:2006sq} & 0.83 &  0.02 & 0.18  \\
RBC/UKQCD '07~\cite{Antonio:2007pb} & 0.720 & 0.013 & 0.037  \\
Aubin, Laiho \& Van de Water '09~\cite{Aubin:2009jh} & 0.724 & 0.008 & 0.028 \\
Average & $0.725 \pm 0.026$ &  &   \\
 \hline \hline
\end{tabular}
\caption{Unquenched lattice QCD determinations of the neutral kaon mixing parameter $\hat{B}_K$.  \textcolor{black}{A plot showing the three $N_f = 2+1$ results and their average is given in Fig.~\ref{fig:BK}.}\label{tab:LQCD_BK}}
\end{center}
\end{table}

The first, by the HPQCD and UKQCD Collaborations~\cite{Gamiz:2006sq}, uses the ``2+1" flavor asqtad-improved staggered gauge configurations\textcolor{black}{~\cite{Susskind:1976jm,Lepage:1998vj,Orginos:1999cr}} generated by the MILC Collaboration~\cite{Bernard:2001av,Bazavov:2009bb}, which include the effects of two degenerate light quarks and one heavier quark with a mass close to that of the physical strange quark.  The calculation also uses staggered valence quarks in the four-fermion operator used to compute $B_K$.  The result for the renormalization-group invariant quantity  $\hat{B}_K$ has a $\sim 22\%$ total uncertainty, which is primarily due to the omission of operators specific to staggered fermions that break flavor symmetry in the lattice-to-continuum operator matching calculation.  Because the other determinations of $B_K$ have much smaller total errors, this result has little impact on the weighted average.

The second calculation by the RBC and UKQCD Collaborations~\cite{Antonio:2007pb} uses 2+1 flavor domain-wall gauge configurations, as well as domain-wall valence quarks in the four-fermion operator used to compute $B_K$.    Because domain-wall quarks have an approximate chiral symmetry~\cite{Kaplan:1992bt,Shamir:1993zy}, it is easier to calculate the renormalization factor needed to determine $B_K$ in the continuum and in the $\bar{\textrm{MS}}$ scheme for domain-wall quarks than for staggered quarks.  Therefore the RBC and UKQCD Collaborations compute the renormalization factor in the RI-MOM scheme nonperturbatively using the method of Rome-Southampton~\cite{Martinelli:1994ty}, and convert it to the $\bar{\textrm{MS}}$ scheme using 1-loop continuum perturbation theory~\cite{Ciuchini:1997bw}.  They estimate that the truncation error due to the perturbative matching is small, $\sim 2\%$.  Although the RBC/UKQCD result is obtained from data at only a single lattice spacing of $a \approx$ 0.11 fm, they include a reasonable $\sim 4\%$ estimate of discretization errors in their total uncertainty, which is based on the scaling behavior of quenched data with the same gluon and valence quark action.   The total error in the RBC/UKQCD calculation of $\hat{B}_K$ is $\sim 5\%$.

Recently, Aubin, Laiho, and Van de Water (ALV) calculated $B_K$ using domain-wall valence quarks on the MILC staggered gauge configurations~\cite{Aubin:2009jh}.  The use of domain-wall valence quarks allows them to compute the renormalization factor nonperturbatively in the RI-MOM scheme just like RBC/UKQCD.  Their result, however, includes a more conservative estimate of the truncation error, $\sim 3\%$, which is based on a comparison with an independent calculation of the renormalization factor using lattice perturbation theory.  Their calculation also improves upon the work of RBC and UKQCD by analyzing data at two lattice spacings, so that they can extrapolate $B_K$ to the continuum limit.   They obtain a total error in $\hat{B}_K$ of $\sim 4\%$.

In order to average the three determinations of $\hat{B}_K$, we must determine which sources of error are correlated between the various calculations.  The HPQCD/UKQCD and ALV calculations both use the staggered gauge configurations generated by the MILC Collaboration.   Aubin, Laiho, and Van de Water, however, use nine independent ensembles of gauge configurations for their analysis, whereas HPQCD/UKQCD only use two ensembles.  Thus the overlap in the two sets of data is quite small, and the statistical errors of the ALV result are sufficiently independent of the statistical errors of the HPQCD/UKQCD result that we treat them as uncorrelated in the average.  The RBC/UKQCD and ALV calculations both use the same 1-loop continuum perturbation theory expression to convert the $B_K$ renormalization factor from the RI-MOM scheme to the $\bar{\textrm{MS}}$ scheme.  Thus we treat the truncation errors as 100\% correlated between the two calculations.  Given these assumptions, we obtain 
\beq
\hat B_K = 0.725 \pm 0.026 
\label{eq:bk}
\eeq
for the weighted average, and we use this value for the unitarity triangle fit presented in Sec.~\ref{sec:SMpredictions}.

%=================================================
\subsection{$B$-meson decay constants and mixing matrix elements}
\label{sec:xi}
%=================================================

The $B$-meson decay constant $f_B$ places a constraint on the CKM unitarity triangle when combined with the experimental branching fraction for $B \to \tau \nu$ leptonic decay.  Because the experimental measurement is difficult, the $B \to \tau \nu$ unitarity triangle constraint is currently quite weak, and is not included in the standard global unitarity triangle fits~\cite{CKMfitter,UTFit}.  Nevertheless, we present the $B$-meson decay constant here with the expectation that the experimental branching fraction will improve and the constraint will be more useful in the future.  Furthermore, we can use the average values for the $B_d$ and $B_s$-meson decays constant, which have smaller errors than any of the individual determinations, to reduce the total uncertainty in the $B_d$ and $B_s$-meson mixing matrix elements, as discussed later in this section.  There have been two 2+1 flavor lattice calculations of the $B$-meson decay constants;  the results for $f_B$ are summarized in the upper panel of Table~\ref{tab:LQCD_fB}, while those for $f_{B_s}$ in the lower panel.

\begin{table}[t]
\begin{center}
\begin{tabular}{lccc}  \hline \hline
& $f_B ({\rm MeV})$ & $\;\;\;\;( \delta f_B)_{\rm stat}$ &  $\;\;\;\;( \delta f_B)_{\rm syst}$ \\
FNAL/MILC '08~\cite{Bernard:2009wr} & 195 & 7 & 9  \\
HPQCD '09~\cite{Gamiz:2009ku} & 190 & 7 & 11  \\
Average & $192.8 \pm 9.9$&   &   \\ \hline\hline
& $f_{B_s} ({\rm MeV})$ & $\;\;\;\;( \delta f_{B_s})_{\rm stat}$ &  $\;\;\;\;( \delta f_{B_s})_{\rm syst}$ \\
FNAL/MILC '08~\cite{Bernard:2009wr} & 243 & 6 & 9  \\
HPQCD '09~\cite{Gamiz:2009ku} & 231 & 5 & 14  \\
Average &  $238.8 \pm 9.5$ &   &   \\  \hline \hline
\end{tabular}
\caption{Unquenched lattice QCD determinations of the $B$-meson decay constants $f_B$ and $f_{B_s}$.  \textcolor{black}{Plots showing the $N_f = 2+1$ results and their averages are given in Figs.~\ref{fig:fB} and~\ref{fig:fBs}.}\label{tab:LQCD_fB}}
\end{center}
\end{table}
The Fermilab Lattice and MILC Collaborations presented preliminary determinations of $f_{B}$ and $f_{B_s}$ at Lattice 2008~\cite{Bernard:2009wr}.  They use the asqtad action for the light $u$, $d$, and $s$-quarks, and the Fermilab action~\cite{ElKhadra:1996mp} for the heavy $b$-quarks.  The largest uncertainty in the decay constants is statistical, and is $\sim$ 3\% for both $f_{B}$ and $f_{B_s}$.  Several systematic uncertainties --- the light-quark discretization error and chiral extrapolation, heavy-quark discretization error, and scale and light-quark mass determination --- all lead to comparable errors of $\sim$ 2\%.

The HPQCD Collaboration recently published a determination of $f_{B}$ and $f_{B_s}$~\cite{Gamiz:2009ku} using staggered light quarks and NRQCD $b$-quarks~\cite{Lepage:1992tx}.  The statistical plus chiral extrapolation errors are comparable to those of Fermilab/MILC.  The largest systematic errors, however, are from the continuum extrapolation ($\sim$ 3\%) and operator matching ($\sim$ 4\%).

Because both decay constant calculations rely upon the MILC gauge configurations, including many overlapping ensembles, we treat the statistical errors as 100\% correlated between the two calculations.  Most of the systematic errors in the two calculations, however, such as those from tuning the quark masses, heavy-quark discretization effects, and operator matching, are independent, so we treat the systematic errors as uncorrelated.  Given these assumptions, we obtain the weighted averages
\bea
f_B & = & 192.8 \pm 9.9 \\
f_{B_s} & = & 238.8 \pm 9.5 .
\eea

In practice, the CKMfitter and UTfit Collaborations do not in fact, use the $B$-meson decay constant to implement the unitarity triangle constraint from $B \to \tau \nu$ decay.  Instead, they construct the ratio ${\rm{B.R.}}(B\to\tau\nu) / \Delta m_d$, where $\Delta m_d$ is the $B_d$-meson oscillation frequency, to reduce the uncertainty from hadronic matrix elements.  The quantity $f_B^2$ cancels in this ratio, such that the ratio depends only on the $B$-meson bag parameter, $B_{B_d}$, which currently has a smaller relative uncertainty than $f_B^2$.  Currently there is only one available 2+1 flavor calculation of the neutral $B$-meson bag parameters by the HPQCD Collaboration~\cite{Gamiz:2009ku}.  They use the same lattice actions and analysis methods as for the decay constants, and obtain
\bea
B_{B_d} & =  &1.26 \pm 0.11 \\
B_{B_s} & =  & 1.33 \pm  0.06 .
\eea
These results are also presented in Table~\ref{tab:LQCD_Bq}.

\begin{table}
\begin{center}
\begin{tabular}{lcc}  \hline \hline
& $\hat{B}_{B_d}$ & $\hat{B}_{B_s}$  \\
HPQCD '09~\cite{Gamiz:2009ku} &$\;\;\;\; 1.26 \pm 0.11\;\;\;\; $ & $1.33 \pm 0.06$ \\ 
 \hline \hline
\end{tabular}
\caption{Unquenched lattice QCD determinations of the neutral $B$-meson bag parameters $\hat{B}_{B_q}$. \label{tab:LQCD_Bq}}
\end{center}
\end{table}

The experimental measurements of the $B_d$- and $B_s$-meson oscillation frequencies, when combined with a calculation of the neutral $B$-meson mixing matrix elements, place additional constraints on the apex of the CKM unitarity triangle.  The weaker of the two constraints comes from $\Delta m_d$, which is proportional to the hadronic matrix element $f_{B_d} \sqrt{\hat{B}_{B_d}}$.  Nevertheless, this constraint plays an important role in the search for new physics because, depending upon the type of physics beyond the Standard Model (BSM) that is present, new physics may affect $B_s$- and $B_d$-mixing independently.  For example, in some minimal flavor violating scenarios, new physics will alter the separate constraints on the apex of the CKM unitarity triangle from $B_s$- and $B_d$-mixing, but not the constraint from their ratio.

Although there has been only one 2+1 flavor calculation of the neutral $B$-meson mixing matrix elements by the HPQCD Collaboration~\cite{Gamiz:2009ku}, there have been two calculations of the decay constant $f_B$, as discussed earlier in this section.  We can therefore use the average values of $f_B$ and $f_{B_s}$ to improve the lattice determinations of the mixing matrix elements $f_{B_d} \sqrt{\hat{B}_{B_d}}$ and $f_{B_s} \sqrt{\hat{B}_{B_s}}$.  We do so by combining the average value of $f_B$ in Table~\ref{tab:LQCD_fB} with the HPQCD determination of $B_{B_d}$ in Table~\ref{tab:LQCD_Bq}.  This procedure reduces the errors in the mixing matrix element to below that from the HPQCD calculation alone, thereby improving the resulting constraint on the unitarity triangle.  We add the errors of $f_B$ and $B_{B_d}$ in quadrature, despite the fact that the average $f_B$ value contains information from the HPQCD decay constant calculation, and is therefore somewhat correlated with the HPQCD $B_{B_d}$ value.  This error treatment is conservative, however, because adding the HPQCD errors for $f_B$ and $B_{B_d}$ in quadrature slightly overestimates the resulting error in $f_{B_d} \sqrt{\hat{B}_{B_d}}$ since correlated statistical fluctuations lead to a smaller error in the product of the two quantities (this is also true for the $B_s$-meson case).  The resulting 2+1 flavor lattice averages for $f_{B_d} \sqrt{\hat{B}_{B_d}}$ and $f_{B_s} \sqrt{\hat{B}_{B_s}}$ are given in Table~\ref{tab:LQCD_fB_Bq}.

\begin{table}
\begin{center}
\begin{tabular}{lcc}  \hline \hline
& $\;\; f_{B} \sqrt{\hat{B}_{B_d}} ({\rm MeV})\;\;$ & $\;\; f_{B_s} \sqrt{\hat{B}_{B_s}} ({\rm MeV}) \;\;$  \\
Average &  $216 \pm 15$ & $275 \pm 13$  \\  \hline \hline
\end{tabular}
\caption{Unquenched lattice QCD averages of the neutral $B$-meson mixing matrix elements $f_{B} \sqrt{\hat{B}_{B_d}}$ and $f_{B_s} \sqrt{\hat{B}_{B_s}}$.  The results are obtained by combining the average decay constants given in Table~\ref{tab:LQCD_fB} with the HPQCD determinations of the bag-parameters presented in Table~\ref{tab:LQCD_Bq}, thereby minimizing the total uncertainties.
\label{tab:LQCD_fB_Bq}}
\end{center}
\end{table}
In practice, the hadronic matrix element $f_{B_d} \sqrt{\hat{B}_{B_d}}$ has larger uncertainties than the corresponding quantity in $B_s$-mixing, $f_{B_s} \sqrt{\hat{B}_{B_s}}$.  This is primarily because current lattice QCD calculations can simulate directly at the physical $s$-quark mass, but must extrapolate to the $u$- and $d$-quark masses.  Therefore the chiral extrapolation error, which is often the dominant systematic, is larger for $f_{B_d} \sqrt{\hat{B}_{B_d}}$ than for $f_{B_s} \sqrt{\hat{B}_{B_s}}$.  In order to minimize the hadronic uncertainty in the $\Delta m_d$ constraint on the unitarity triangle, we therefore replace $f_{B_d} \sqrt{\hat{B}_{B_d}}$ with $  f_{B_s} \sqrt{\hat{B}_{B_s}}/\xi$, where $\xi \equiv  f_{B_s} \sqrt{\hat{B}_{B_s}}/f_{B_d} \sqrt{\hat{B}_{B_d}}$ is an $SU(3)$-breaking ratio that can currently be determined more accurately in lattice calculations than the individual matrix elements, as discussed below.

The more stringent neutral $B$-meson mixing constraint on the unitarity triangle comes from the ratio of the oscillation frequencies, $\Delta m_s / \Delta m_d$, because many uncertainties are reduced in the lattice calculation of $\xi$. There have been two recent 2+1 flavor lattice QCD calculations of $\xi$;  the results are summarized in Table~\ref{tab:LQCD_xi}.

\begin{table}
\begin{center}
\begin{tabular}{lccc} \hline \hline
& $\xi$ &  $\;\;\;\;(\delta \xi)_{\rm stat}$ &  $\;\;\;\;(\delta \xi)_{\rm syst}$ \\
 FNAL/MILC '08~\cite{Evans:2009} & 1.205 & 0.036 & 0.037  \\
HPQCD '09~\cite{Gamiz:2009ku} & 1.258 & 0.025 & 0.021  \\
Average & $1.243 \pm 0.028$ &   &   \\ \hline \hline
\end{tabular}
\caption{Unquenched lattice QCD determinations of the $SU(3)$-breaking ratio $\xi$. \textcolor{black}{A plot showing the two $N_f = 2+1$ results and their average is given in Fig.~\ref{fig:xi}.} \label{tab:LQCD_xi}}
\end{center}
\end{table}
The Fermilab Lattice and MILC Collaborations presented a preliminary calculation of $\xi$ at Lattice 2008~\cite{Evans:2009}, while the HPQCD Collaboration has recently published a determination of $\xi$ to $2.6\%$~\cite{Gamiz:2009ku} accuracy.  Both groups use the same lattice actions and methods as they did for the $B$-meson decay constant calculations.  The largest uncertainties in Fermilab-MILC's determination of $\xi$ are from statistics and from the chiral-continuum extrapolation, both of which are $\sim$ 3\%.  They obtain a total error in $\xi$ of $\sim 4\%$.   HPQCD's largest source of uncertainty is also statistics and the chiral-continuum extrapolation, which together contribute $\sim 2\%$ to the total uncertainty.   Because both calculations of $\xi$ rely on the MILC gauge configurations, in taking the average of the two results, we treat the statistical errors as 100\% correlated.   We treat the systematic errors as uncorrelated because most of the systematic errors are independent since they use different heavy-quark actions and operator renormalization methods.  Given these assumptions, we obtain the weighted average
\beq
\xi = 1.243 \pm 0.028 
\eeq
for use in the unitarity triangle analysis.

%=================================================
\subsection{$|V_{ub}|$}
\label{sec:Vub}
%=================================================

The CKM matrix element $|V_{ub}|$ also places a constraint on the apex of the CKM unitarity triangle.  It can be determined by combining experimental measurements of the branching fraction for semileptonic $B \to \pi \ell \nu$ decay with lattice QCD calculations of the $B \to \pi \ell \nu$ form factor.  There have been two exclusive determinations of $|V_{ub}|$ based on 2+1 flavor lattice calculations;  the results are summarized in the upper panel of Table~\ref{tab:LQCD_Vqb}.

\begin{table}
\begin{center}
\begin{tabular}{lccc} \hline\hline
 & $\left|V_{ub}\right| \times 10^{-3} $ & $\;\;\;\;( \delta V_{ub})_{\rm exp} $ &  $\;\;\;\;( \delta V_{ub})_{\rm theo}$ \\
HPQCD '06~\cite{Dalgic:2006dt} + HFAG Winter '09~\cite{HFAG_FPCP_09} &  3.40 & 0.20 & $^{+0.59}_{-0.39}$ \\
FNAL/MILC '08~\cite{Bailey:2008wp} + BABAR '06~\cite{Aubert:2006px}  &  3.38 & $\sim$ 0.20 & $\sim$  0.29 \\
Average & $3.42 \pm 0.37$ &    &    \\ 
\hline\hline
 & $\left|V_{cb}\right| \times 10^{-3} $ & $\;\;\;\;( \delta V_{cb})_{\rm exp} $ &  $\;\;\;\;( \delta V_{cb})_{\rm theo}$ \\
$B\to D \ell \nu$:  FNAL/MILC  '04~\cite{Okamoto:2004xg} + HFAG Winter  '09~\cite{HFAG_FPCP_09} & 39.1 & 1.4 & 0.9  \\
$B\to D^* \ell \nu$: FNAL/MILC  '08~\cite{Bernard:2008dn} + HFAG Winter '09~\cite{HFAG_FPCP_09} & 38.3 & 0.5 & 1.0  \\
Average & $38.6 \pm 1.2$ &   &  \\
\hline\hline
\end{tabular}
\caption{Exclusive determinations of the CKM matrix elements $|V_{ub}|$ and $|V_{cb}|$ from unquenched lattice QCD calculations. \textcolor{black}{Plots showing the $N_f = 2+1$ results and their averages are given in Figs.~\ref{fig:Vcb} and~\ref{fig:Vub}.} \label{tab:LQCD_Vqb}}
\end{center}
\end{table}

In 2006 the HPQCD Collaboration published the first unquenched computation of the $B \to \pi \ell \nu$ semileptonic form factor using asqtad valence light quarks and NRQCD valence b-quarks~\cite{Dalgic:2006dt} and
MILC 2+1 flavor dynamical gauge configurations. The $B \to \pi \ell \nu$ form factor is more difficult to compute numerically than other lattice quantities such as $B_K$ or $\xi$, and consequently has a larger total error.  Because of the poor statistics associated with lattice data at nonzero momentum, the largest source of uncertainty in the HPQCD form factor calculation is the 10\% statistical plus chiral extrapolation error.  When the HPQCD result for the form factor is combined with the latest Heavy Flavor Averaging Group (HFAG) average for the $B \to \pi \ell \nu$ branching fraction~\cite{HFAG_FPCP_09}, one obtains $|V_{ub}|$ with a total error of $\sim 16\%$, only $\sim 6\%$ of which comes from the experimental uncertainty in the branching fraction.

The Fermilab Lattice and MILC Collaborations recently published an improved determination of the $B \to \pi \ell \nu$ semileptonic form factor and $|V_{ub}|$ using staggered light quarks and Fermilab $b$-quarks~\cite{Bailey:2008wp}.  As in the case of the HPQCD calculation, the largest source of uncertainty is statistics plus chiral-continuum extrapolation, which leads to a $\sim 6\%$ error in the form factor.  Fermilab/MILC, however, extract $|V_{ub}|$ in a different manner than HFAG.  They perform a simultaneous fit to the lattice data and the 12-bin BABAR experimental data~\cite{Aubert:2006px} using a fit function based on analyticity and crossing symmetry~\cite{Bourrely:1980gp,Boyd:1994tt,Lellouch:1995yv,Boyd:1997qw}, leaving the relative normalization between lattice and experiment as a free parameter to be determined in the fit.  With this method they reduce the total uncertainty in $|V_{ub}|$ by combining the lattice and experimental information in an optimal, model-independent manner.  They obtain a total error in $|V_{ub}|$ of $\sim 11\%$.

Because the HPQCD and Fermilab/MILC calculations both use the MILC gauge configurations, we treat their statistical errors as 100\% correlated when taking the average.  We also treat the experimental errors in the two determinations of $|V_{ub}|$ as 100\% correlated.  This is a conservative assumption because the HPQCD extraction comes from the HFAG average, which is obtained from many experimental measurements of the branching fraction including the 12-bin BABAR analysis.  We treat the systematic errors in the two calculations as uncorrelated, since that they use different actions for the heavy quarks and different methods for the lattice-to-continuum operator matching.  Given these assumptions, we obtain
\beq
|V_{ub}| = ( 3.42 \pm 0.37) \times 10^{-3} 
\label{vubexclAV}
\eeq
for the weighted average. \textcolor{black}{Note that in our averaging procedure we symmetrize the HPQCD systematic error. For this reason the central value of the average (\ref{vubexclAV}) is slightly larger than both the HPQCD and Fermilab/MILC central values.}

%=================================================
\subsection{$|V_{cb}|$}
\label{sec:Vcb}
%=================================================

The CKM matrix element $|V_{cb}|$ normalizes the base of the CKM unitarity triangle.  Therefore it implicitly enters many of the constraints on the apex of the CKM unitarity triangle, including those coming from $B_K$ and $|V_{ub}|$.
$|V_{cb}|$ can be determined by combining experimental measurements of the branching fractions for $B \to D \ell \nu$ or $B \to D^* \ell \nu$ semileptonic decay, in combination with lattice QCD calculations of the relevant form factor at zero recoil.  There have been two exclusive determinations of $|V_{cb}|$ based on 2+1 flavor lattice calculations;  the results are summarized in the lower panel of Table~\ref{tab:LQCD_Vqb}.

The Fermilab Lattice and MILC Collaborations presented the first unquenched lattice determination of the $B \to D \ell \nu$ form factor at Lattice 2004~\cite{Okamoto:2004xg}.  Like other Fermilab/MILC calculations of heavy-light meson quantities, this calculation uses staggered light quarks and Fermilab $b$- and $c$-quarks.  Because the $B \to D \ell \nu$ form factor at zero recoil can be obtained by a carefully constructed double ratio of matrix elements, it can be computed very precisely from lattice calculations.  The statistical error in the resulting form factor determination is $\sim 2\%$, and all systematic errors are $\sim 1\%$ or less.  Although the result is obtained from data at only a single lattice spacing, Fermilab/MILC include an estimate of discretization effects in their error budget.  When the result for the $B \to D \ell \nu$ form factor is combined with the latest HFAG average of the experimental branching fraction~\cite{HFAG_FPCP_09}, it leads to a determination of $|V_{cb}|$ with a $\sim 4\%$ total error, of which $\sim 2\%$ is lattice theoretical and $\sim 4\%$ is experimental.

More recently, the Fermilab Lattice and MILC Collaborations published the first unquenched lattice determination of the $B \to D^* \ell \nu$ form factor~\cite{Bernard:2008dn}.  This calculation also uses staggered light quarks and Fermilab $b$- and $c$-quarks, and obtains the form factor at zero recoil from a double ratio of matrix elements.  It improves upon the earlier determination of the $B \to D \ell \nu$ form factor, however, by using data at three lattice spacings, and performing a more sophisticated chiral-continuum extrapolation.  When the result for the $B \to D^* \ell \nu$ form factor is combined with the latest HFAG average of the experimental branching fraction~\cite{HFAG_FPCP_09}, it leads to a determination of $|V_{cb}|$ with a $\sim 3\%$ total error, of which $\sim 2.5\%$ is lattice theoretical and $\sim 1.5\%$ is experimental. We note that the average of $B\to D^*$ data yields a poor chi-square per degree of freedom ($\chi^2_{\rm min}/{d.o.f.} = 39/21$). For this reason, we rescale the experimental error quoted in Table~\ref{tab:LQCD_Vqb} by a factor $\sim1.4$.

Because the Fermilab/MILC calculations are computed on some of the same ensembles and using the same lattice actions and methods, we treat the theoretical errors as 100\% correlated between the extractions of $|V_{cb}|$ from $D$ and $D^*$ final states. We assume that the experimental errors are independent between the two measurements. Given these assumptions we obtain the weighted average
\beq
|V_{cb}|_{\rm excl} = ( 38.6 \pm 1.2) \times 10^{-3} \; 
\label{vcbexclusive}
\eeq
for use in the unitarity triangle fit.

%=================================================
\subsection{$\kappa_\varepsilon$}
\label{sec:kappaepsilon}
%=================================================
Buras and Guadagnoli~\cite{Buras:2008nn} have pointed out that corrections to $\varepsilon_K$ that had typically been neglected in the unitarity triangle analysis due to the large errors on $B_K$ and $|V_{cb}|$ \textcolor{black}{(with the exception of Refs.~\cite{Andriyash:2003ym,Andriyash:2005ax})} are actually substantial, and amount to a $\sim 8\%$ correction to the Standard Model prediction for $\varepsilon_K$.  In this section, we \textcolor{black}{provide the first estimate of the correction factor $\kappa_\varepsilon$ using a 2+1 flavor lattice QCD calculation of ${\rm Im}  \left[A (K\to \pi \pi(I=2))\right]$~\cite{Li:2008kc}.  We follow the notation of Ref.~\cite{Buras:2008nn}.  Our result supersedes previous estimates of $\kappa_\varepsilon$~\cite{Bardeen:1986vz,Anikeev:2001rk,Buras:2003zz,Blanke:2007wr}, and is in perfect agreement with the recent determination of Ref.~\cite{Buras:2008nn}.}

It is conventional to define $K\to\pi\pi$ matrix elements in terms of definite isospin amplitudes by
\bea A(K^0\to\pi\pi(I))=A_I e^{i\delta_I},
\eea
\bea A(\overline{K}^0\to\pi\pi(I))=-A^*_I e^{i\delta_I}.
\eea
CP-violation in the kaon system is then parameterized in terms of
\bea\label{eq:epsilon_K} \varepsilon_K = e^{i\phi_\varepsilon}\sin{\phi_\varepsilon}\left(\frac{\textrm{Im}(M^K_{12})}{\Delta M_K}+P_0 \right),
\eea
and
\bea \varepsilon'_K = \frac{ie^{i(\delta_2-\delta_0)}}{\sqrt{2}}\omega[P_2-P_0],
\eea
where
\bea\label{eq:P_12} P_0 \equiv \frac{\textrm{Im} A_0}{\textrm{Re} A_0}, \ \ \  P_2 \equiv \frac{\textrm{Im}A_2}{\textrm{Re}A_2}, \ \ \ \omega\equiv \frac{\textrm{Re}A_2}{\textrm{Re}A_0}.
\eea
The first term in parenthesis in Eq.~(\ref{eq:epsilon_K}) is the short distance contribution to kaon mixing; it is the part that is conventionally normalized by the bag-parameter $B_K$.  The second term, $P_0$, is due to long distance contributions to kaon mixing, and is related to the ratio of $K\to\pi\pi$ decay amplitudes in the $\Delta I=1/2$ channel defined in Eq.~(\ref{eq:P_12}).
In the usual unitarity triangle analysis, $\phi_\varepsilon$ is taken to be $\pi$/4, and $P_0$ is taken to be negligible compared to the first term in parenthesis in Eq.~(\ref{eq:epsilon_K}).  However, these corrections are not negligible in the current analysis, and the corrections coming from $\phi_\varepsilon \neq \pi/4$ and $P_0\neq 0$ are small but in the same direction \cite{Buras:2008nn}.  

Following Ref.~\cite{Buras:2008nn}, we define an overall multiplicative correction factor for $\varepsilon_K$ that accounts for $\phi_\varepsilon \neq \pi/4$ and $P_0\neq 0$,
\bea\label{eq:kappa_epsilon} \kappa_\varepsilon = \sqrt{2} \sin{\phi_\varepsilon} \overline{\kappa}_\varepsilon,
\eea
where $\overline{\kappa}_\varepsilon$ parameterizes the correction to $|\varepsilon_K|$ coming from $P_0$.  To good approximation, we have
\bea \overline{\kappa}_\varepsilon \sim 1 + \frac{P_0}{\sqrt{2}|\varepsilon_K|},
\eea
and given that
\bea \textrm{Re}\left(\varepsilon'_K/\varepsilon_K\right)\sim \frac{\omega}{\sqrt{2}|\varepsilon_K|}(P_2-P_0),
\eea
we see that
\bea\label{eq:kappabar_epsilon} \overline{\kappa}_\varepsilon = 1-\frac{1}{\omega}\textrm{Re}(\varepsilon'_K/\varepsilon_K) + \frac{P_2}{\sqrt{2}|\varepsilon_K|}.
\eea

All of the quantities entering $\kappa_\varepsilon$, Eqs.~(\ref{eq:kappa_epsilon}) and (\ref{eq:kappabar_epsilon}), are well determined from experiment, assuming the Standard Model, except for $P_2$.  In Eq.~(\ref{eq:P_12}), we need theory input for $\textrm{Im} A_2$ to determine $P_2$.  The amplitude $\textrm{Im}A_2$ is non-vanishing due to the electroweak penguin contribution to $K\to\pi\pi$ decays.  We use the result of lattice calculations of this amplitude in our analysis.  There is only one $2+1$ flavor result for this quantity with rather large systematic errors associated with the use of leading order chiral perturbation theory \cite{Li:2008kc}, obtained by the RBC/UKQCD Collaborations.  Their value is given in Table~\ref{tab:ImA2unquench}, where the first error is statistical and the second is due to the systematic error in the determination of the leading order low energy constant of chiral perturbation theory ($\chi$PT), and the truncation of $\chi$PT to tree-level.  Although the quoted error budget does not include typical lattice errors such as finite volume effects, scale setting, and discretization errors, they are almost certainly much smaller than the errors attributed to the use of leading order chiral perturbation theory.  For comparison, we also mention lattice results for $\textrm{Im}A_2$ in the quenched approximation, collected in Table~\ref{tab:ImA2}.  Some of these quenched calculations are more thorough than others at assessing systematic errors.  However, all of their systematic error budgets are necessarily incomplete, due to the uncontrolled nature of the quenched approximation, and we do not attempt to estimate this error here.  We take a simple average of the five quenched lattice results and note that it is very similar to the $2+1$ flavor result.  We use only the $2+1$ flavor result for our determination of $\kappa_\epsilon$ for use in the unitarity triangle fits.  

\begin{table}[t]
\begin{center}
\begin{tabular}{lcc} \hline\hline
2+1 Flavor & \ $\textrm{Im}A_2 \times 10^{13}$ GeV &  \\[0.5mm] 
RBC/UKQCD '08~\cite{Li:2008kc} & $-7.9\pm1.6 \pm3.9$    &\\ \hline\hline
\end{tabular}
\caption{2+1 flavor lattice value for $\textrm{Im}A_2$.  Errors are statistical and systematic, respectively.  \textcolor{black}{A plot comparing the $N_f = 2+1$ result with several quenched determinations is given in Fig.~\ref{fig:ImA2}.}\label{tab:ImA2unquench}}
\end{center}
\end{table}
\begin{table}[t]
\begin{center}
\begin{tabular}{lcc} \hline\hline
Quenched & \ $\textrm{Im}A_2 \times 10^{13}$ GeV &  \\[0.5mm] 
RBC '01~\cite{Blum:2001xb} & $-12.6$  \\
CP-PACS '01~\cite{Noaki:2001un} & $-9.1$   \\
${\rm SPQ_{CD}R}$ '04~\cite{Boucaud:2004aa} & $-5.5$ \\
Babich et al '06~\cite{Babich:2006bh} & $-9.2$\\
Yamazaki '08~\cite{Yamazaki:2008hg} & $-11.8$  \\
Average & $-9.6$    &\\ \hline\hline
\end{tabular}
\caption{Quenched lattice values for $\textrm{Im}A_2$. \label{tab:ImA2}}
\end{center}
\end{table}
\begin{table}[t]
\begin{center}
\begin{tabular}{lcl}
\hline\hline
$\phi_\varepsilon=(43.51\pm0.05)^{\circ}$ & \;\;\;\;\; &
$|\varepsilon_K|=(2.229\pm0.012)\times 10^{-3}$  \\
$\omega=0.0450$ & &
$\textrm{Re}(\varepsilon'_K/\varepsilon_K)=1.68\pm0.19\times 10^{-3}$  \\
$\textrm{Re}A_2 = 1.50\times 10^{-8} \; {\rm GeV}$ & &
$\textrm{Im}A_2 = (-7.9\pm 4.2)\times 10^{-13}\; {\rm GeV}$
 \vphantom{\Big(} \\
\hline
\hline
\end{tabular}
\caption{Inputs used to determine $\kappa_\varepsilon$.\label{tab:kappa_eps}}\end{center}
\end{table}
All inputs used to determine $\kappa_\varepsilon$ are given in Table~\ref{tab:kappa_eps}.  We take the most recent experimental world average for $\textrm{Re}(\varepsilon'/\varepsilon)$~\cite{Blucher2009}, noting that we inflate the errors according to the PDG prescription because of the somewhat low confidence level ($13\%$) in the world average.  Using these values in Eqs.~(\ref{eq:P_12}), (\ref{eq:kappa_epsilon}), and (\ref{eq:kappabar_epsilon}) we find 
\bea \kappa_\varepsilon=0.92\pm0.01,
\eea
in agreement with Ref.~\cite{Buras:2008nn}.  The $50\%$ error in the $2+1$ flavor determination of $\textrm{Im} A_2$ dominates the error in $\kappa_\epsilon$.  We note for comparison that if we use the average quenched value of $\textrm{Im} A_2$, assigning to it a $100\%$ error, we find $\kappa_\epsilon=0.92\pm0.02$. 

%Using the experimental value of $\phi_\varepsilon = (43.51\pm 0.05)^{\circ}$, we have $\kappa_\varepsilon=0.974\overline{\kappa}_\varepsilon$.

%=================================================
\subsection{$f_K$}
\label{sec:fk}
%=================================================

The kaon decay constant $f_K$ enters the CKM unitarity triangle through $\varepsilon_K$. 
\textcolor{black}{Because experiments can only measure the product $f_K \times |V_{us}|$, lattice calculations are needed to obtain $f_K$ by itself.}  There have already been four 2+1 flavor lattice QCD determinations of $f_K$ using different valence and sea quark actions, and several more calculations are underway.  Thus $f_K$ is one of the best-known hadronic weak matrix elements.  Table~\ref{tab:LQCD_fK} summarizes the current status of 2+1 flavor lattice QCD calculations of $f_K$. 

\begin{table}[t]
\begin{center}
\begin{tabular}{lccc} \hline\hline
& $f_K ({\rm MeV})$ & $\;\;\;\;(\delta f_K)_{\rm stat}$ & $\;\;\;\;(\delta f_K)_{\rm syst}$ \\
MILC '07~\cite{Bernard:2007ps} & 156.5 & 0.4 & $^{+1.0}_{-2.7}$  \\
HPQCD/UKQCD '07~\cite{Follana:2007uv} & 157 &  1 & 2  \\
RBC/UKQCD '08~\cite{Allton:2008pn} & 149.6 & 3.6 & 6.3  \\
Aubin, Laiho \& Van de Water '08~\cite{Aubin:2008ie} & 153.9 & 1.7 & 4.4 \\
Average & $155.8 \pm 1.7$  &  &\\
\hline\hline
\end{tabular}
\caption{Unquenched lattice QCD determinations of the kaon decay constant $f_K$. \textcolor{black}{A plot showing the four $N_f = 2+1$ results and their average is given in Fig.~\ref{fig:fK}.} \label{tab:LQCD_fK}}
\end{center}
\end{table}

The MILC Collaboration published the first 2+1 flavor determination of $f_K$ in 2004~\cite{Aubin:2004fs}, and updated the result at Lattice 2007 by including data with lighter quarks and finer lattice spacings~\cite{Bernard:2007ps}.  The largest source of uncertainty in their calculation is from the extrapolation to the physical light quark masses and the continuum.  A small but non-negligible error also arises due to the determination of the absolute lattice scale needed to convert dimensionful quantities into physical units.  MILC first determines the relative scale $r_1/a$ from the heavy-quark potential.  Next they obtain the absolute scale $r_1 = 0.3108(15)(^{+26}_{-79})$ by tuning $f_\pi$ to be equal to the experimental value.   The uncertainty in $r_1$ leads to an uncertainty in $f_K$ of $^{+0.25}_{-0.75}$, which is 25\% of the total error.  The remaining finite volume effects and EM effects are an order of magnitude or more smaller, and the total uncertainty in the MILC Collaboration's determination of $f_K$ is  $\sim 2\%$.

The HPQCD Collaboration published a determination of $f_K$ using a mixed-action method with highly-improved staggered quarks~\cite{Follana:2006rc} on the MILC asqtad-improved staggered gauge configurations~\cite{Follana:2007uv}.  The largest source of uncertainty in their calculation is from the determination of the scale.  They use the MILC Collaboration's determination of the relative scale $r_1/a$ to convert dimensionful quantities from lattice units to $r_1$ units.  They obtain the value $r_1 = 0.321(5)$, independently, however, using the $\Upsilon$ spectrum computed with nonrelativistic $b$-quarks on the MILC ensembles~\cite{Gray:2005ur}. The uncertainty in $r_1$ leads to an uncertainty in $f_K$ of 1.1\%.  The remaining statistical and systematic errors are all much smaller, and the total error in $f_K$  is 1.3\%.  

The RBC and UKQCD Collaborations published an independent determination of $f_K$ using domain-wall quarks~\cite{Allton:2008pn}.  Because they obtain their result using only a single lattice spacing of $a \approx 0.11$ fm, the dominant uncertainty in their result is from discretization errors.  They estimate these errors to be 6\% using power-counting arguments.  Because the remaining statistical and systematic errors are all much smaller, the total error in $f_K$ is 6.3\%.

Aubin, Laiho, and Van de Water presented a preliminary determination of $f_K$ using a mixed-action method with domain-wall valence quarks on the MILC staggered gauge configurations at Lattice 2008~\cite{Aubin:2008ie}.  The largest source of uncertainty in their calculation is from the chiral and continuum extrapolation, which they estimate to be 2.3\%.  The 1.6\% error from the uncertainty in the scale $r_1$ however, is close in size, so the total error in $f_K$ is 3.0\%.

Because the HPQCD, MILC and ALV calculations all use the MILC gauge configurations, we treat the statistical errors as 100\% correlated when taking the average.  Because ALV use the MILC Collaboration's determination of the scale $r_1$ from $f_\pi$, the scale uncertainties are also 100\% correlated between the calculations.  We take the scale uncertainty in HPQCD's calculation to be uncorrelated, however, because they use a largely independent determination of $r_1$ based on the $\Upsilon$ spectrum.  We also treat the remaining systematic errors as uncorrelated between the HPQCD, MILC, and ALV calculations because they use different valence quark formulations.  The calculation of the RBC and UKQCD Collaborations is independent of the other results, and we therefore take the errors to be completely uncorrelated.  Given these assumptions, we obtain the weighted average
\beq
f_K = (155.8 \pm 1.7) \; {\rm MeV}
\eeq
to be used in the unitarity triangle analysis presented in Sec.~\ref{sec:SMpredictions}.

%=================================================
\section{Other inputs to the fit of the unitarity triangle}
\label{sec:OtherInputs}
%=================================================

Table~\ref{tab:utinputs} summarizes the set of inputs that we use in the fit.  We obtain $\alpha$ from the isospin analysis of $B\to (\pi\pi,\rho\rho,\rho\pi)$ decays (the description of the method we use can be found in Refs.~\cite{Gronau:1990ka,Snyder:1993mx,Quinn:2000by} and the experimental inputs are taken from Ref.~\cite{HFAG_Moriond_09}). We take the direct determination of $\gamma$ from the model-independent UTfit analysis of $B\to D^{(*)} K^{(*)}$ decays~\cite{Bona:2005vz,Bona:2006ah} (the experimental inputs used are taken from Ref.~\cite{HFAG_Moriond_09}). The inclusive determination of $|V_{cb}|$ deviates by more than 2 $\sigma$ from the average of exclusive results that we quote in Eq.~(\ref{vcbexclusive}). For use in the unitarity triangle fit we combine the three determinations of $|V_{cb}|$ from inclusive and exclusive ($D$ and $D^*$) modes. Taking into account the correlations between the errors of the two exclusive determinations of $|V_{cb}|$ and assuming no correlation between inclusive and exclusive analyses, we obtain:
\bea
\left|V_{cb}\right|_{\rm excl + incl} = \left( 40.3 \pm 1.0 \right) \times 10^{-3} \; ,
\eea 
where the error has been appropriately rescaled following the PDG prescription. We quote the inclusive determination of $|V_{ub}|$ from the most recent GGOU analysis~\cite{Gambino:2007rp,HFAG_FPCP_09}.  Because, however, the extraction of $|V_{ub}|_{\rm incl}$ depends strongly on the theoretical framework adopted~\cite{HFAG_FPCP_09}, we adopt a conservative stance and omit $|V_{ub}|_{\rm incl}$ from the set of measurements that we include in the full unitarity triangle fit. Our predictions for the Standard Model parameters in the following section are independent of $|V_{ub}|$, and our conclusions regarding indications of new physics in Sec.~\ref{sec:NewPhys} are relatively insensitive to the value of $|V_{ub}|$. \textcolor{black}{Apart from the inputs listed in Table~\ref{tab:utinputs}, we take $G_F$, $m_K$, $m_W$, $m_{B_d}$ and $m_{B_s}$ from the Particle Data Group~\cite{Yao:2006px}.}
\begin{table}[t]
\begin{center}
\begin{tabular}{ll}
\hline\hline
$\left| V_{cb} \right|_{\rm incl} =
(41.31 \pm 0.76) \times 10^{-3}$~\cite{HFAG_FPCP_09} $\;\;\;$&
$\left| V_{ub} \right|_{\rm incl} = 
(40.3 \pm 1.5^{+2.0}_ {-2.5}) \times 10^{-4} $~\cite{HFAG_FPCP_09}  \cr
$\Delta m_{B_d} = (0.507 \pm 0.005)\; {\rm ps}^{-1}$~\cite{HFAG_PDG_09}  & 
$\Delta m_{B_s} = (17.77 \pm 0.10 \pm 0.07)\;  {\rm ps}^{-1}$~\cite{Evans:2007hq}  \vphantom{\Big(} \\
$\alpha = (89.5 \pm 4.3)^{\rm o}$& 
$\gamma = (78 \pm 12)^{\rm o}$~\cite{Bona:2005vz,Bona:2006ah}
 \vphantom{\Big(} \\
$\eta_1 = 1.51 \pm 0.24$~\cite{Herrlich:1993yv} & 
$m_{t, pole} = (172.4 \pm 1.2) \; {\rm GeV}$~\cite{:2008vn} \vphantom{\Big(}\\
$\eta_2 = 0.5765 \pm 0.0065$~\cite{Buras:1990fn}  & 
$m_c(m_c) = (1.268 \pm 0.009 ) \; {\rm GeV}$~\cite{Allison:2008xk}\vphantom{\Big(}\\
$\eta_3 = 0.47 \pm 0.04$~\cite{Herrlich:1995hh}  &  
$\varepsilon_K = (2.229 \pm 0.012 ) \times 10^{-3}$~\cite{Yao:2006px} \vphantom{\Big(} \\
$\eta_B = 0.551 \pm 0.007$~\cite{Buchalla:1996ys} &
$\lambda = 0.2255  \pm 0.0007$~\cite{Antonelli:2008jg}\vphantom{\Big(}  \\ 
$S_{\psi K_S} = 0.672 \pm 0.024$~\cite{HFAG_Moriond_09}  &
 \vphantom{\Big(} \\
\hline
\hline
\end{tabular}
\caption{Inputs used in the unitarity triangle fit.  \textcolor{black}{Note that the most precise determination of $m_c$ is  obtained from lattice QCD~\cite{Allison:2008xk}.}  \label{tab:utinputs}}
\end{center}
\end{table}

%=================================================
\section{Standard Model Predictions}
\label{sec:SMpredictions}
%=================================================
 
In this section we extract the Standard Model predictions for $\hat B_K$, $|V_{cb}|$ and $|V_{ub}/V_{cb}|$.  We use only the three constraints from $S_{\psi K_S}$, $\Delta M_{B_s}/\Delta M_{B_d}$ and $\varepsilon_K$, and do not include the constraints from $|V_{ub}|$, $\alpha$ and $\gamma$ in the fit because predictions are almost completely insensitive to their impact.  The analytical formulae for $\varepsilon_K$ and $\Delta M_{B_s}/\Delta M_{B_d}$ can be found, for instance, in Ref.~\cite{Lunghi:2009sm}. 

\bigskip

We obtain the prediction for $\hat B_K$ by excluding the direct lattice determination of $\hat B_K$ from the chi-square. The dominant source of uncertainty in the extraction of $\hat B_K$ stems from the strong dependence of $\varepsilon_K$ on $|V_{cb}|$ ($\varepsilon_K  \propto |V_{cb}|^4$). This issue is even more problematic because of the discrepancy between the extraction of $|V_{cb}|$ from exclusive and inclusive decays. For this reason we perform the analysis both with and without the inclusive determination of $|V_{cb}|$. Note that when combining the inclusive and exclusive extractions of $|V_{cb}|$ we follow the PDG prescription for inflating the error when combining inconsistent measurements.  We find:
\bea
( \hat B_K )_{\rm fit}  =  
\begin{cases}
1.09 \pm 0.12 & \left|V_{cb} \right|_{\rm excl} \cr 
0.903 \pm 0.086 &\left|V_{cb} \right|_{\rm incl} \cr 
0.98 \pm 0.10 & \left|V_{cb} \right|_{\rm excl+incl} \cr 
\end{cases}
\eea
The comparison of these predictions with the lattice determination of $\hat B_K$ given in Eq.~(\ref{eq:bk}) yields a deviation at the $2.9\sigma$, $2\sigma$ and $2.4\sigma$ level, respectively.
\begin{figure}[t]
\begin{center}
\includegraphics[width= 0.48\linewidth]{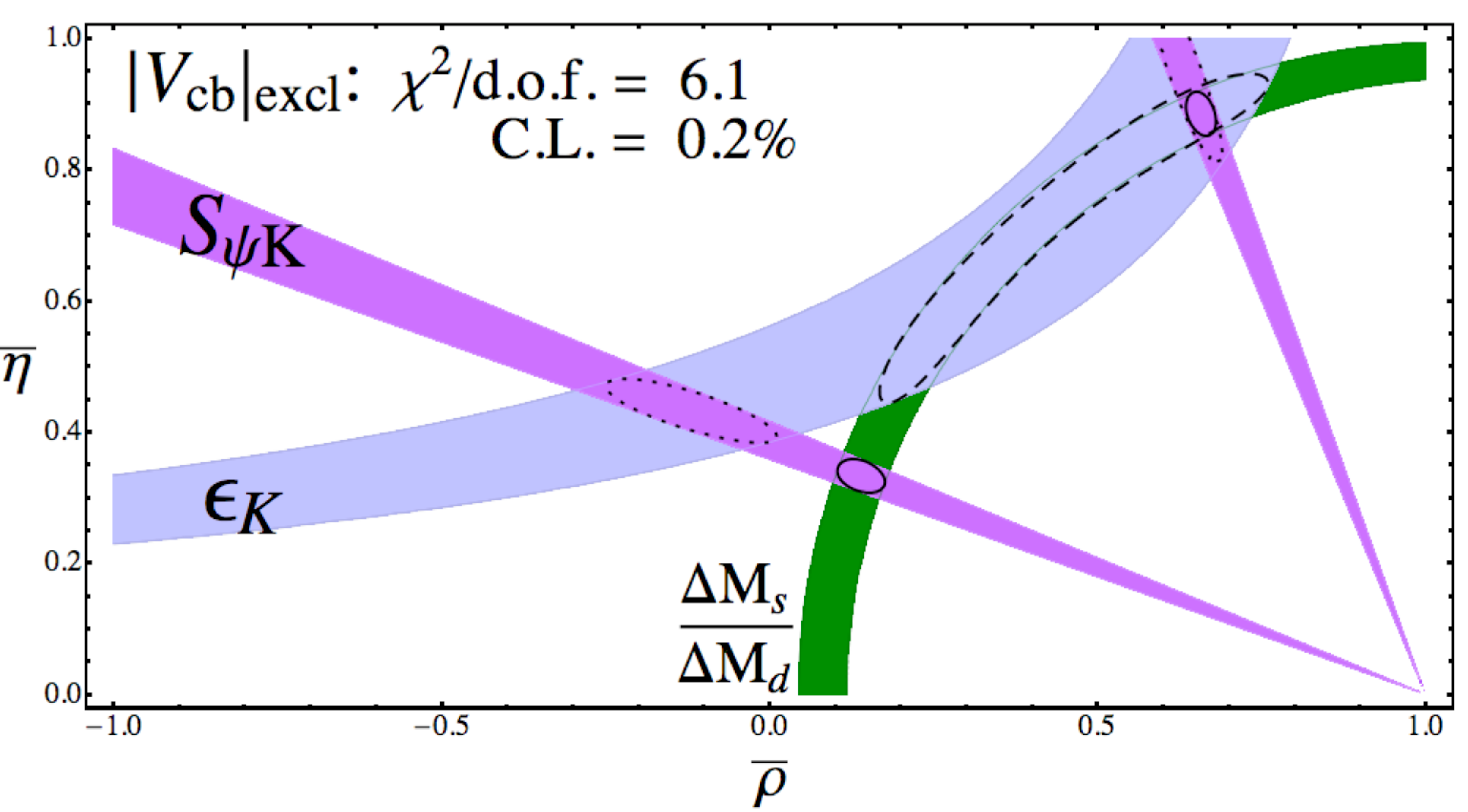}
\includegraphics[width= 0.48\linewidth]{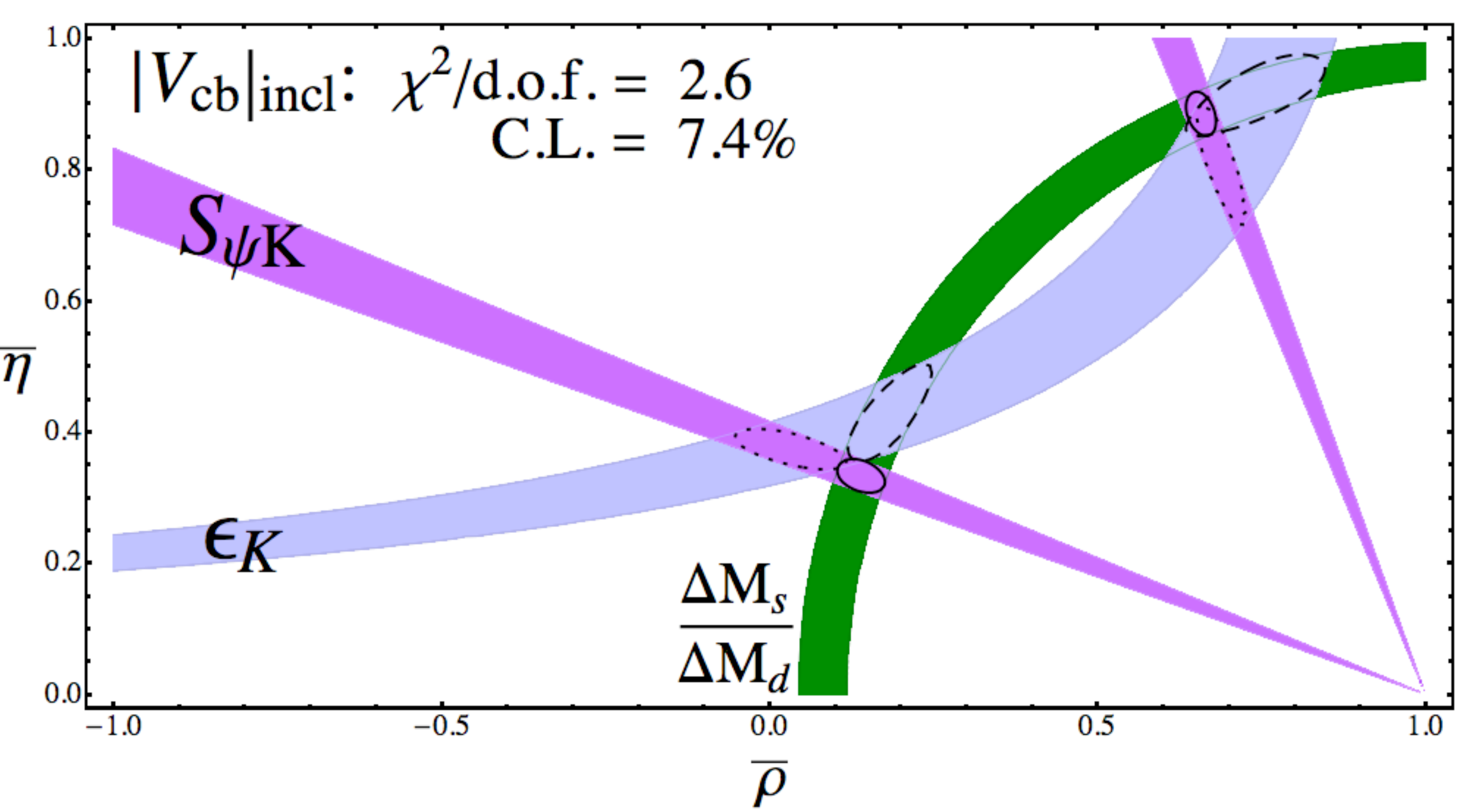}
\caption{Impact of $\varepsilon_K$ on the UT fit. The solid, dashed and dotted contours are obtained by omitting $\varepsilon_K$,  $S_{\psi K}$ and $\Delta M_{B_s}/\Delta M_{B_d}$, respectively. The left and right panels use the exclusive and inclusive $|V_{cb}|$ determinations, respectively. \label{fig:ut}}
\end{center}
\end{figure}
We obtain the prediction for $|V_{cb}|$ in a similar fashion by excluding the inclusive and exclusive determinations of $|V_{cb}|$ from the chi-square.  We find:
\bea
\left| V_{cb} \right|_{\rm fit} = (43.0 \pm 0.9) \times 10^{-3} \; .
\eea
This prediction deviates by $3.0\sigma$ and $1.3\sigma$ from the exclusive and inclusive determinations of $|V_{cb}|$, respectively.  

Figure~\ref{fig:ut} illustrates the (2 -- 3)$\sigma$ tension in the fit to the unitarity triangle. In the left and right panels we use $|V_{cb}|$ from exclusive and inclusive semileptonic $B$ decays, respectively. The solid, dashed and dotted contours are obtained by omitting $\varepsilon_K$,  $S_{\psi K}$ and $\Delta M_{B_s}/\Delta M_{B_d}$, respectively. They correspond to three scenarios in which new physics affects $K$ mixing, the phase and the amplitude of $B_d$ mixing. An alternative measure of the tension in the UT fit is the minimum $\chi^2$ per degree of freedom: when we include all three constraints we obtain $\chi^2_{\rm min}/{\rm dof} = 6.1 \; (2.6)$ using $|V_{cb}|_{\rm excl \; (incl)}$, corresponding to a confidence level of 0.2\% (7.4\%).

\begin{figure}[t]
\begin{center}
\includegraphics[width= 0.48\linewidth]{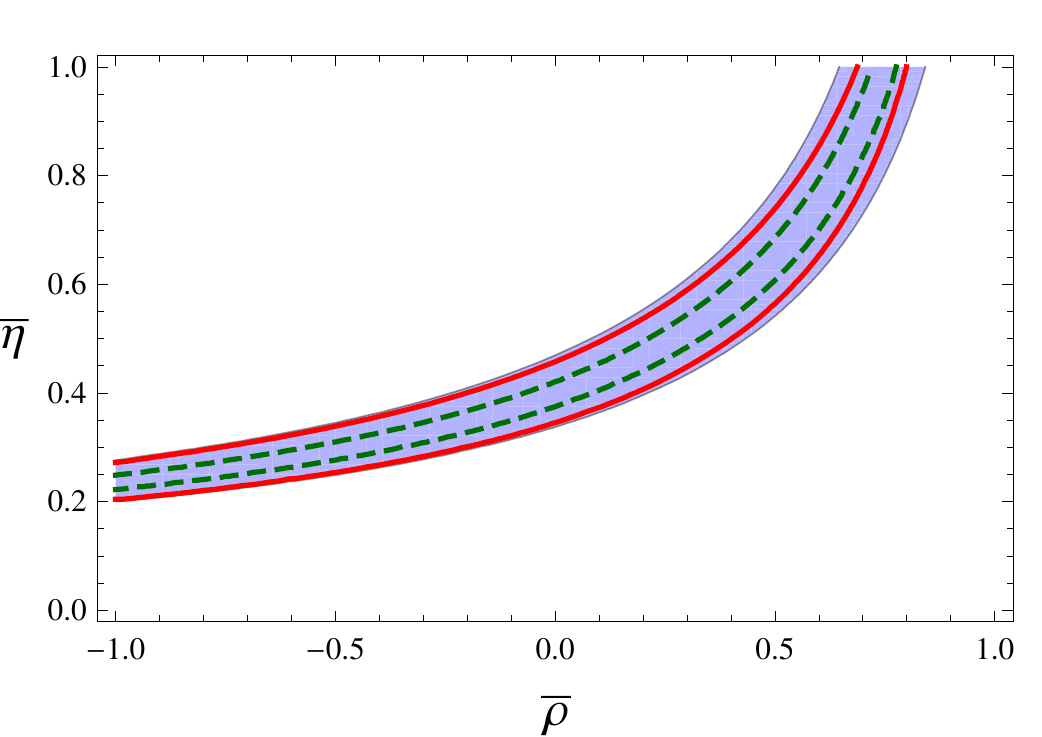}
\caption{Impact of $|V_{cb}|$ (solid red line) and $\hat B_K$ (dashed green line) on the $\varepsilon_K$ error band. The uncertainties induced by variations of $m_c$, $m_t$, $\eta_i$ and $\kappa_\varepsilon$ have a negligible impact on the $\varepsilon_K$ error budget.
\label{fig:ek}}
\end{center}
\end{figure}

It is interesting to note that the errors on the fitted values of $\hat B_K$ ($\sim 10\%$ -- $12\%$) are much larger than the corresponding lattice uncertainty ($\sim 3.5 \%$); therefore, improvements on the latter are not poised to have a sizable effect on the UT tension. In contrast, the errors on the direct and indirect determinations of $|V_{cb}|$ are similar (about $2\%$), indicating that improvements on the theoretical predictions for exclusive and inclusive semileptonic $B$ decays will have a huge impact on our understanding of this issue. This discussion is summarized in Fig.~\ref{fig:ek}, which shows the relative impact of the present  $|V_{cb}|_{\rm excl}$ and $\hat B_K$ uncertainties on the total $\varepsilon_K$ error band. 

Finally, we note that the SM prediction for the ratio $|V_{ub}/V_{cb}|$ is in good agreement with the lattice expectation and deviates by only $1.6\sigma$ from the inclusive ratio:
\bea
\left| \frac{V_{ub}}{V_{cb}} \right|  =  
\begin{cases}
0.0846 \pm 0.0035 & \rm fit \cr 
0.089 \pm 0.010 & \rm exclusive \cr
0.0969 \pm 0.0068 & \rm inclusive \cr
\end{cases} \; .
\eea
%
%=================================================
\section{Interpretation as New Physics}
\label{sec:NewPhys}
%=================================================

In this section we assume that physics beyond the Standard Model does not affect tree-level processes at the current level of precision, and that any sign of new physics must arise due to higher-order loop effects.  Given these assumptions, it is well known~\cite{Lunghi:2007ak, Buras:2008nn, Lunghi:2008aa, Buras:2009pj} that the $\sim 2 \sigma$ tension in the fit to the unitarity triangle can be interpreted as a manifestation of new physics effects in $\varepsilon_K$ and/or $B_d$ mixing.  In order to test the consistency of these two hypotheses with the current measurements, we describe the two new physics possibilities using the following model-independent parametrization:
\bea
\varepsilon_K & = & C_\varepsilon \left( \varepsilon_K\right)_{\rm SM}  \; ,  \label{ekpar} \\
M_{12}^{d} & = & r_d^2 \;  e^{i 2 \theta_d}  \left( M_{12}^{d} \right)_{\rm SM}\; , \label{Bdpar}
\eea
where $M_{12}^d$ is the matrix element of the complete effective Hamiltonian between $B^0$ and $\bar B_0$ states.  A value of $C_\varepsilon \neq 1$ would move the location of the $\varepsilon_K$ band, while the presence of a $r_d \neq 1$ and a non-vanishing $\phi_d$ would alter the following three unitarity triangle constraints:
\bea
\Delta M_{B_d} & = & \Delta M_{B_d}^{\rm SM} \; r_d^2  \; , \\
\beta_{\rm eff} & = & \beta + \theta_d \; , \\
\alpha_{\rm eff} & = & \alpha - \theta_d \; ,
\eea
where $\beta_{\rm eff}$ and $\alpha_{\rm eff}$ are the angles extracted from the CP asymmetries in $B\to J/\psi K$ and $B\to (\pi\pi,\rho\rho,\rho\pi)$, respectively. \textcolor{black}{Although the presence of new physics effects in $B_s$ mixing is a very interesting possibility, we do not consider it in the model-independent analysis presented in this work for the following reason. Because of the smallness of the phase of the CKM element $V_{ts}$, any evidence of CP violation in $B_s$ decays translates immediately into evidence for physics beyond the SM. The golden mode studied at CDF and D0 is the time dependent CP asymmetry in the decay $B_s \to J/\psi \phi$, which presently deviates by about $2\sigma$ from the SM expectation ($\sim 0$). Unfortunately, however, a new phase in $B_s$ mixing does not affect any of the observables that enter the unitarity triangle fit. Therefore, signs of new physics in $B_d /K$ and $B_s$ mixing can be connected only in specific new physics scenarios in which new phases in $K$, $B_d$ and $B_s$ mixing have a common origin (see for instance Refs.~\cite{Buras:2008nn,Lunghi:2009sm} for a discussion of this point).}

\begin{figure}[t]
\begin{center}
\includegraphics[width= 0.48\linewidth]{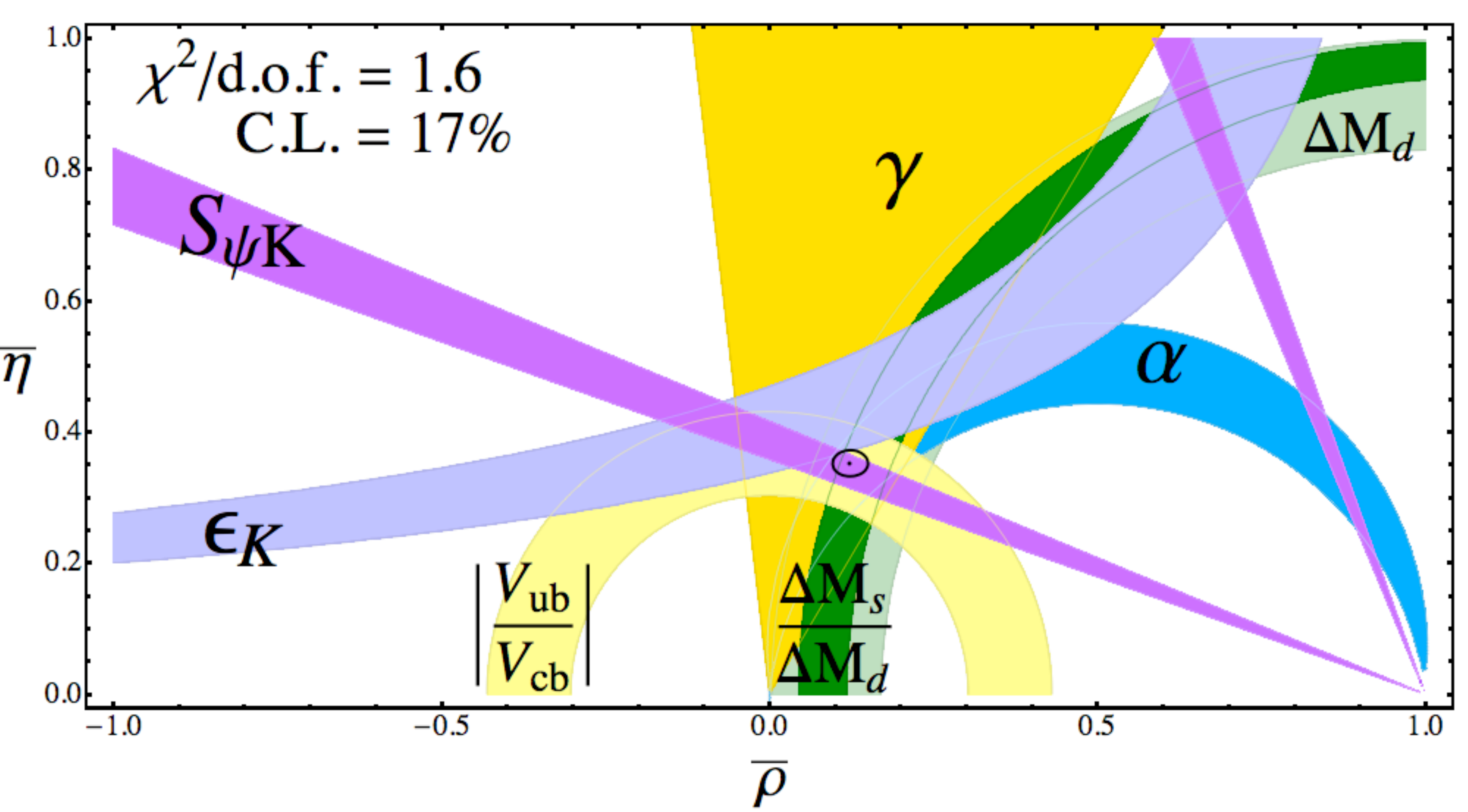}
\caption{Fit to the unitarity triangle in the SM. We average the inclusive and exclusive determinations of $|V_{cb}|$, but use only the exclusive determination of  $|V_{ub}|$. The black contour is obtained from the minimization of the complete chi-squared.  
\label{fig:utfit-tot}}
\end{center}
\end{figure}

From the inspection of Fig.~\ref{fig:ut} we see that, as long as we consider only $\varepsilon_K$, $\Delta M_{B_s}/\Delta M_{B_d}$, $S_{\psi K_S}$ and $|V_{cb}|$, the data are not able distinguish between scenarios with new physics in the $K$ or $B_d$ sectors. In fact, both the solid (new physics in $K$ mixing) and dashed (new physics in $B_d$ mixing) contours in Fig.~\ref{fig:ut} have $\chi^2_{\rm min} \simeq 0$. This tie, however, is broken by the inclusion of constraints on $|V_{ub}|$ (from exclusive semileptonic $b\to u$ decays), $\alpha$ (from $B\to (\pi\pi, \rho\rho,\rho\pi)$ decays) and $\gamma$ (from $B\to D^{(*)} K^{(*)}$ decays). Figure~\ref{fig:utfit-tot} shows the resulting full fit to the unitarity triangle using the combined inclusive + exclusive determination of $|V_{cb}|$.

\begin{figure}[t]
\begin{center}
\includegraphics[width= 0.48\linewidth]{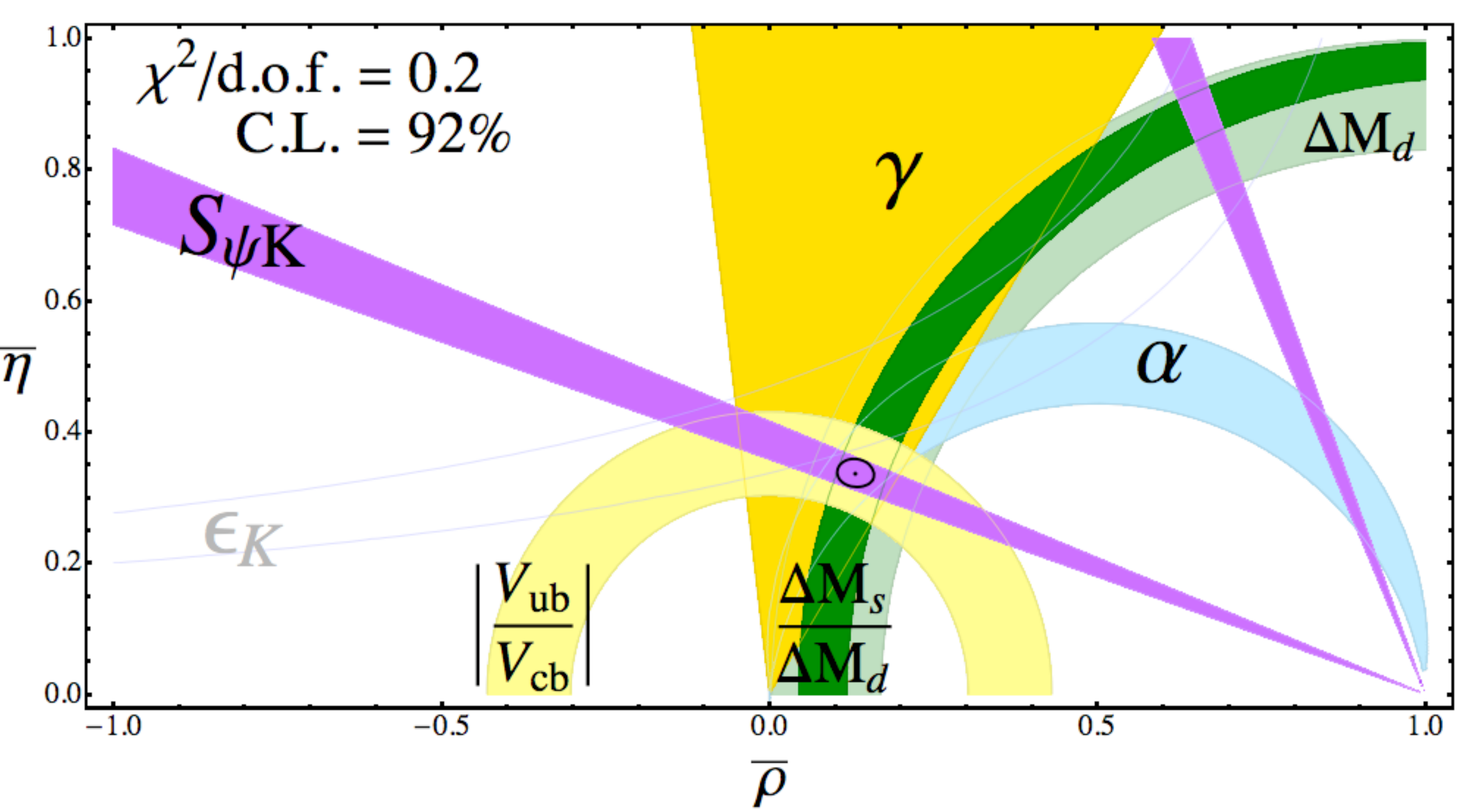}
\includegraphics[width= 0.48\linewidth]{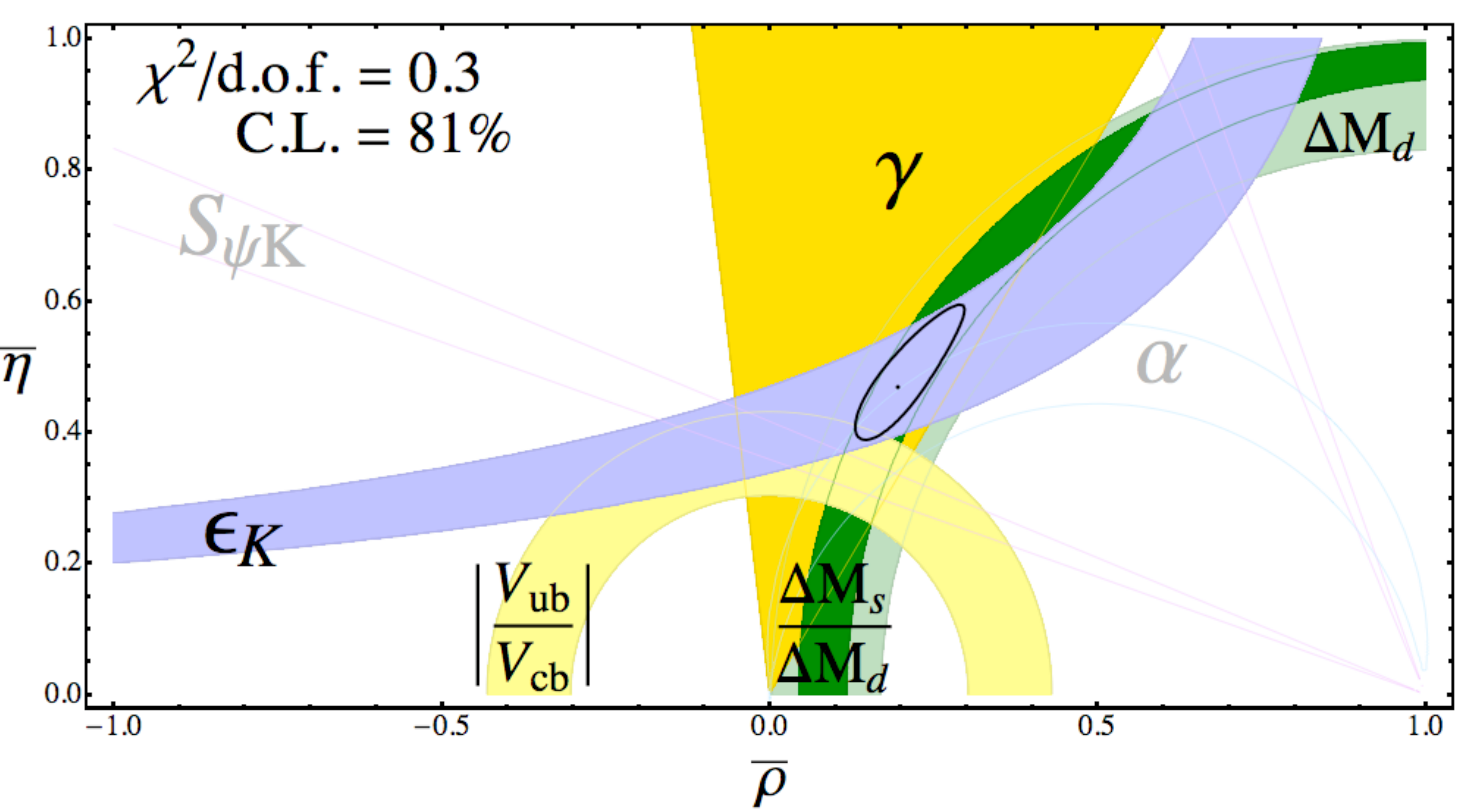}
\includegraphics[width= 0.48\linewidth]{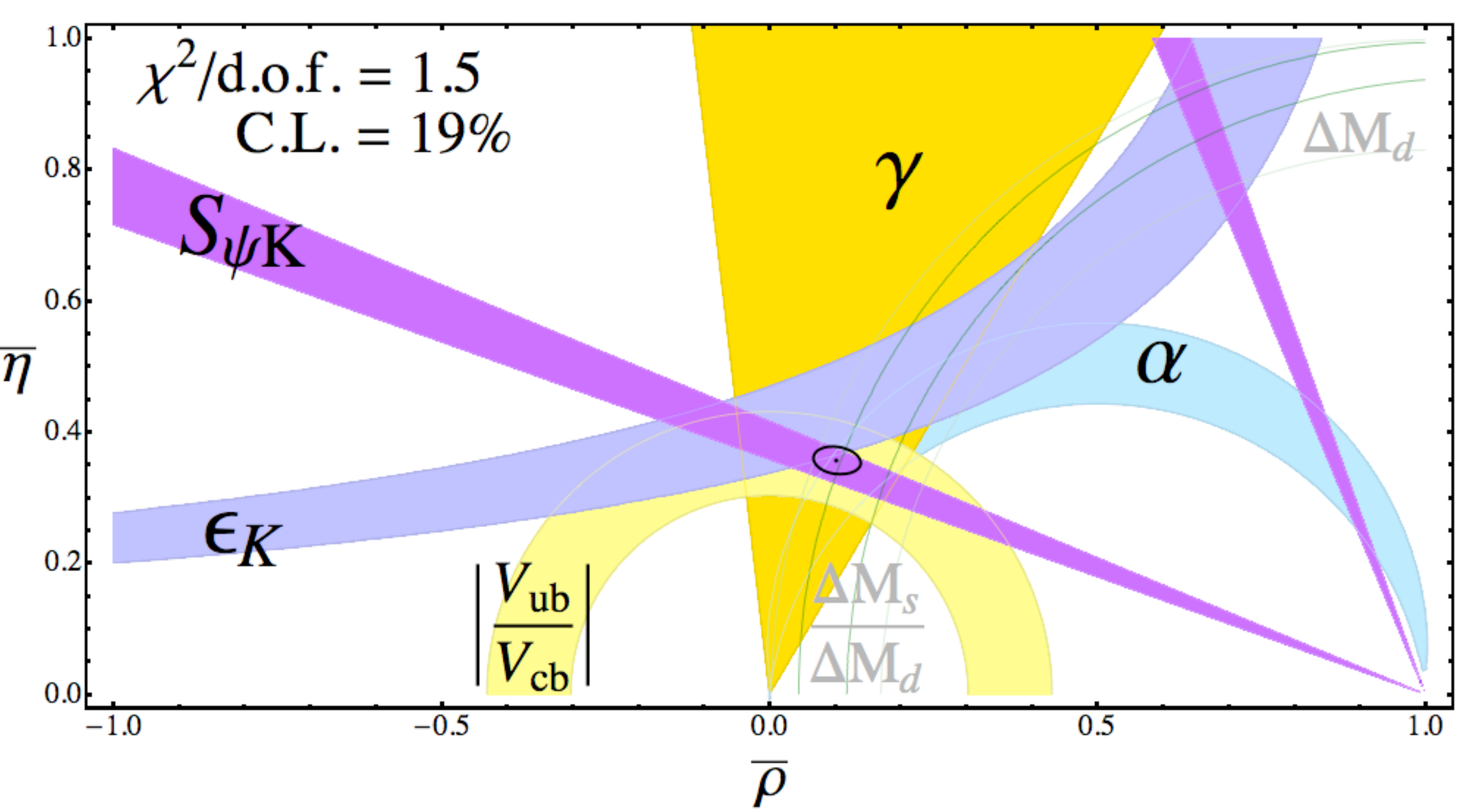}
\caption{Full fit to the unitarity triangle. Upper left panel: The black contour is obtained without the inclusion of the $\varepsilon_K$ constraint. Upper right panel: The black contour is obtained without the inclusion of the $\alpha$ and $\beta$ constraints. Lower panel: The black contour is obtained without the inclusion of the $\Delta M_{B_d}$ \textcolor{black}{ and $\Delta M_{B_s}/\Delta M_{B_d}$ constraints.}\label{fig:utNP}}
\end{center}
\end{figure}

In order to test the hypothesis that new physics only affects neutral kaon-mixing, we minimize the chi-square while excluding $\varepsilon_K$ from the fit.  The solid contour in the left panel of Fig.~\ref{fig:utNP} shows the allowed $(\bar\rho,\bar\eta)$ region in this scenario.  Adopting the parametrization in Eq.~(\ref{ekpar}) we obtain the following value for the new physics contribution to $\varepsilon_K$:
\bea
( C_\varepsilon)_{\rm fit}  =  
\begin{cases}
1.47 \pm 0.17 & \left|V_{cb} \right|_{\rm excl} \; ,  \cr 
1.21 \pm 0.11 &\left|V_{cb} \right|_{\rm incl} \; , \cr 
1.32 \pm 0.14 & \left|V_{cb} \right|_{\rm excl+incl}\; . \cr 
\end{cases}
\label{ce}
\eea
In the upper right and lower panels of Fig.~\ref{fig:utNP} we consider scenarios in which only new physics in $B_d$ mixing is allowed. For the sake of simplicity we consider only two extreme cases in which we take $(\theta_d\neq 0, r_d=1)$ and $(\theta_d=0,r_d \neq 1)$. In the former case $S_{\psi K}$ and the extraction of $\alpha$ are affected by new physics contributions and must be excluded from the fit; in the latter one, only $\Delta M_{B_s}/\Delta M_{B_d}$ receives contributions. The fitted values of the new physics parameters $\theta_d$ and $r_d$ are:
\bea
( \theta_d)_{\rm fit}  =  
\begin{cases}
(-4.3 \pm 2.1)^{\rm o}  & \left|V_{cb} \right|_{\rm excl}  \cr  
(-2.8 \pm 1.9)^{\rm o} &\left|V_{cb} \right|_{\rm incl}  \cr 
(-3.4 \pm 2.0)^{\rm o} & \left|V_{cb} \right|_{\rm excl+incl}\cr 
\end{cases}
\quad {\rm and} \quad
( r_d)_{\rm fit}  =  
\begin{cases}
(0.940 \pm 0.036)  & \left|V_{cb} \right|_{\rm excl}  \cr 
(0.950 \pm 0.036)&\left|V_{cb} \right|_{\rm incl}  \cr 
(0.946 \pm 0.036) & \left|V_{cb} \right|_{\rm excl+incl} \; .\cr 
\end{cases} 
\label{td}
\eea
In this case, the tension between $\Delta M_{B_s}/\Delta M_{B_d}$, $\varepsilon_K$ and $|V_{ub}|$ reduces the quality of the fit: the fit omitting the constraint from $\varepsilon_K$ has a confidence level of 91\%, while the fit omitting the constraints from $S_{\psi K}$ and $\alpha$ and from $\Delta M_{B_s}/\Delta M_{B_d}$ have a confidence level of 23\% and 30\%, respectively. Thus the scenario with new physics in $K$ mixing is favored by present data. This can also be seen from the inspection of Eqs.~(\ref{ce}) and (\ref{td}): $C_\varepsilon$ deviates from $C_\varepsilon^{\rm SM} = 1$ at a higher confidence level than $\theta_d$ from $\theta_d^{\rm SM} = 0$. 

\textcolor{black}{Finally, it should be noted that the marked preference for new physics in the kaon sector is a direct consequence of our inclusion of the lower determination of $|V_{ub}|$ from exclusive semileptonic decays in the  fit. As can be seen from the upper right-hand plot in Fig.~\ref{fig:utNP}, further removal of the $|V_{ub}/V_{cb}|$ constraint results in a  fit with a high confidence level (CL=81\%).  The overlap of the constraints from $\varepsilon_K$ and $\Delta M_s / \Delta M_d$, however, corresponds to a very large value of $|V_{ub}/V_{cb}| = 0.120 \pm 0.017$. The correlation between possible new physics in $B_d$ mixing and a large implied value of $|V_{ub}/V_{cb}|$ is well known, and has been discussed in Refs.~\cite{Buras:2008nn,Altmannshofer:2009ne,Buras:2009if}.}

%

%=================================================
\section{Conclusions}
\label{sec:Conc}
%=================================================

Lattice QCD calculations that include the effects of the dynamical up, down, and strange quarks are becoming standard, and allow reliable calculations of hadronic weak matrix elements with all sources of uncertainty under control.  Because there are now multiple lattice calculations of most of the hadronic matrix elements that enter the unitarity triangle fit, it is essential to average these results in order to reduce the theoretical uncertainties and obtain the most sensitive test of new physics in the flavor sector possible.  We have therefore presented averages for the hadronic weak matrix elements that enter the standard global fit of the CKM unitarity triangle.  Although we do not know the precise correlations between different lattice calculations of the same quantity, we have accounted for correlations between the different lattice results in a conservative manner in our averages.  Whenever there is any correlation between the statistical or a particular systematic error in different lattice calculations, we assume that the degree of correlation is 100\%.    Our lattice averages of hadronic weak matrix elements are therefore appropriate for use in phenomenological analyses such as the global CKM unitarity triangle fit.

When these up-to-date lattice averages of the hadronic weak matrix elements are used in a global fit of the CKM unitarity triangle, we find a (2--3)$\sigma$ tension.  \textcolor{black}{As was first pointed out by Lunghi and Soni~\cite{Lunghi:2008aa}, this tension is primarily between the three most precise constraints on the unitarity triangle from $\sin(2 \beta)$, $\Delta M_{B_s} / \Delta M_{B_d}$, and $\varepsilon_K$, and is largely independent of the value of $|V_{ub}|$, which differs significantly between determinations using inclusive and exclusive semileptonic decays.  We confirm their observation and put it on an even stronger footing by using lattice averages that include more recent lattice calculations and take into account correlations.}  The significance of the tension depends upon whether we use the exclusive or inclusive determination of $|V_{cb}|$, which disagree by $\sim 2 \sigma$.  If we assume that new physics does not affect tree-level processes at the current level of precision, this tension can be interpreted as a sign of new physics either in neutral kaon mixing or in neutral $B$-meson mixing.  We find that the current data prefer the scenario in which the new physics is in kaon mixing;  this can be seen by the fact that the confidence level of the global fit increases significantly when we remove the constraint from the $\epsilon_K$ band leaving all others unchanged.  The tension between the $\epsilon_K$ band and the other constraints is enhanced by our inclusion of the correction factor $\kappa_\epsilon$, which lowers the Standard Model prediction for $\epsilon_K$ by 8\%. \textcolor{black}{This factor has been recently included by the UTfit collaboration~\cite{Bona:2009ze} (they find a similar tension in the fit), but not yet by the CKMfitter group~\cite{CKMfitter}.}

The errors in the hadronic weak matrix elements needed as inputs to the unitarity triangle analysis will continue to decrease over the next few years, as the results included in these averages are updated and as new independent results using different lattice actions from other collaborations appear.   If the tension observed in the current global fit persists as the theoretical errors are reduced, this may indeed be a sign of new physics.  This will be difficult to ascertain conclusively, however, unless the inclusive and exclusive determinations of $|V_{cb}|$ converge.  Thus a better understanding of the theoretical errors in both determinations is a high priority for flavor physics.  Lattice QCD calculations of weak matrix elements are truly living up to their promise and may ultimately lead to the discovery of new physics in the quark flavor sector.

%=================================================
\section*{Acknowledgments}
%=================================================

\textcolor{black}{We thank Andreas Kronfeld and Steve Gottlieb for entertaining comments on the manuscript. We also acknowledge interesting dicussions with Diego Guadagnoli, Andrzej Buras and Amarjit Soni.}

\appendix
%=================================================
\section{Summary plots of lattice QCD averages}
\label{sec:App}
%=================================================

In this section we provide summary plots of the lattice QCD averages discussed in Sec.~\ref{sec:LatticeInputs}.

\begin{figure}[t]
\begin{center}
\includegraphics[width=0.6\linewidth]{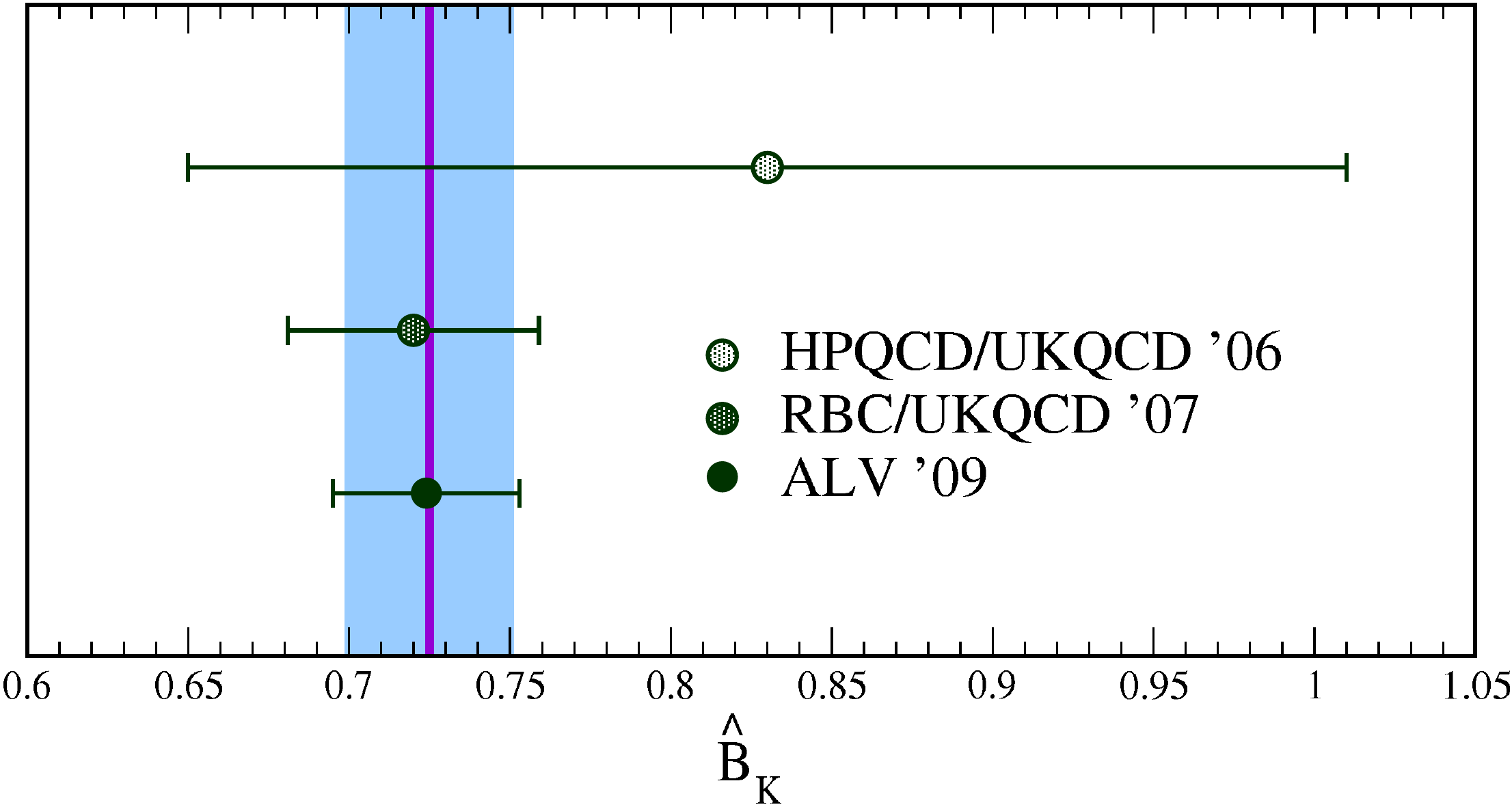}
\caption{Unquenched lattice average of the neutral kaon mixing parameter $\hat{B}_K$.
The three $N_f = 2+1$ lattice inputs are given by shaded green circles with error bars, while the resulting average is denoted as a purple vertical line with a blue error band.
\label{fig:BK}}
\end{center}
\end{figure}

\begin{figure}[t]
\begin{center}
\includegraphics[width=0.6\linewidth]{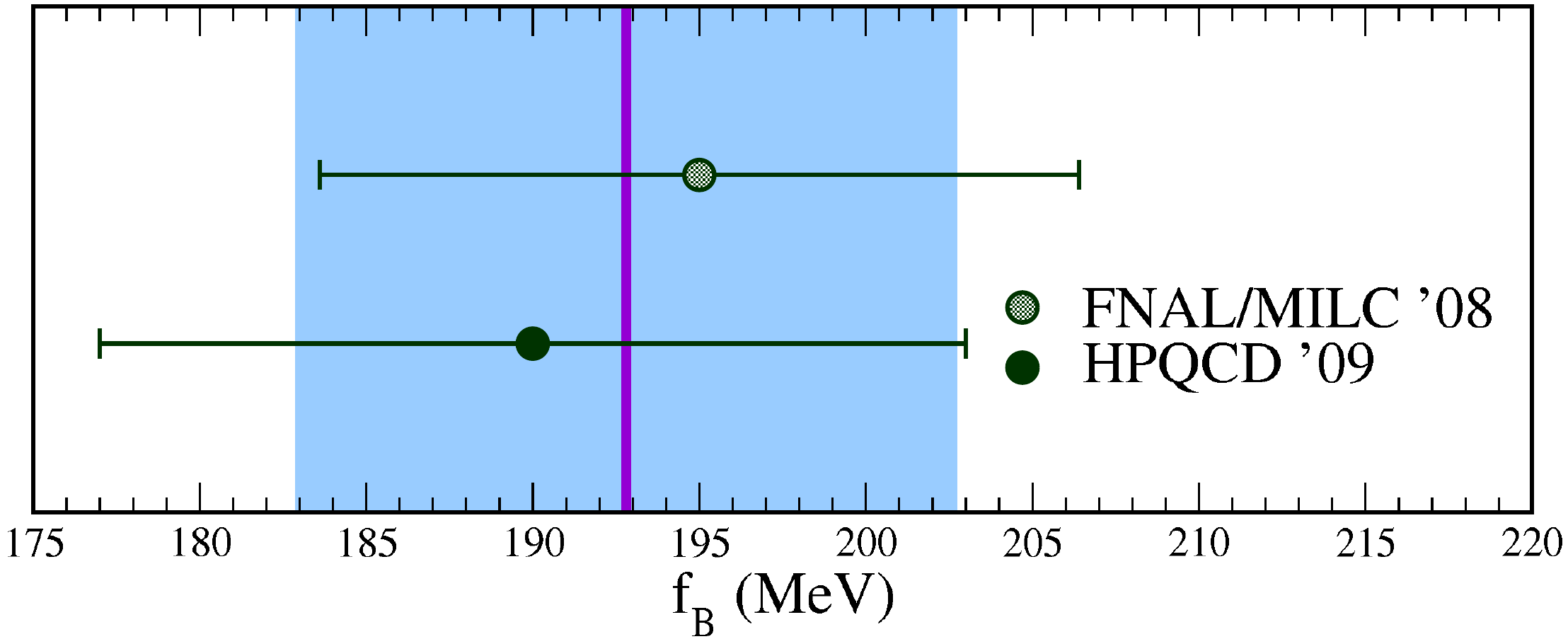}
\caption{Unquenched lattice average of the leptonic decay constant $f_B$.
The two $N_f = 2+1$ lattice inputs are given by shaded green circles with error bars, while the resulting average is denoted as a purple vertical line with a blue error band.
\label{fig:fB}}
\end{center}
\end{figure}

\begin{figure}[t]
\begin{center}
\includegraphics[width=0.6\linewidth]{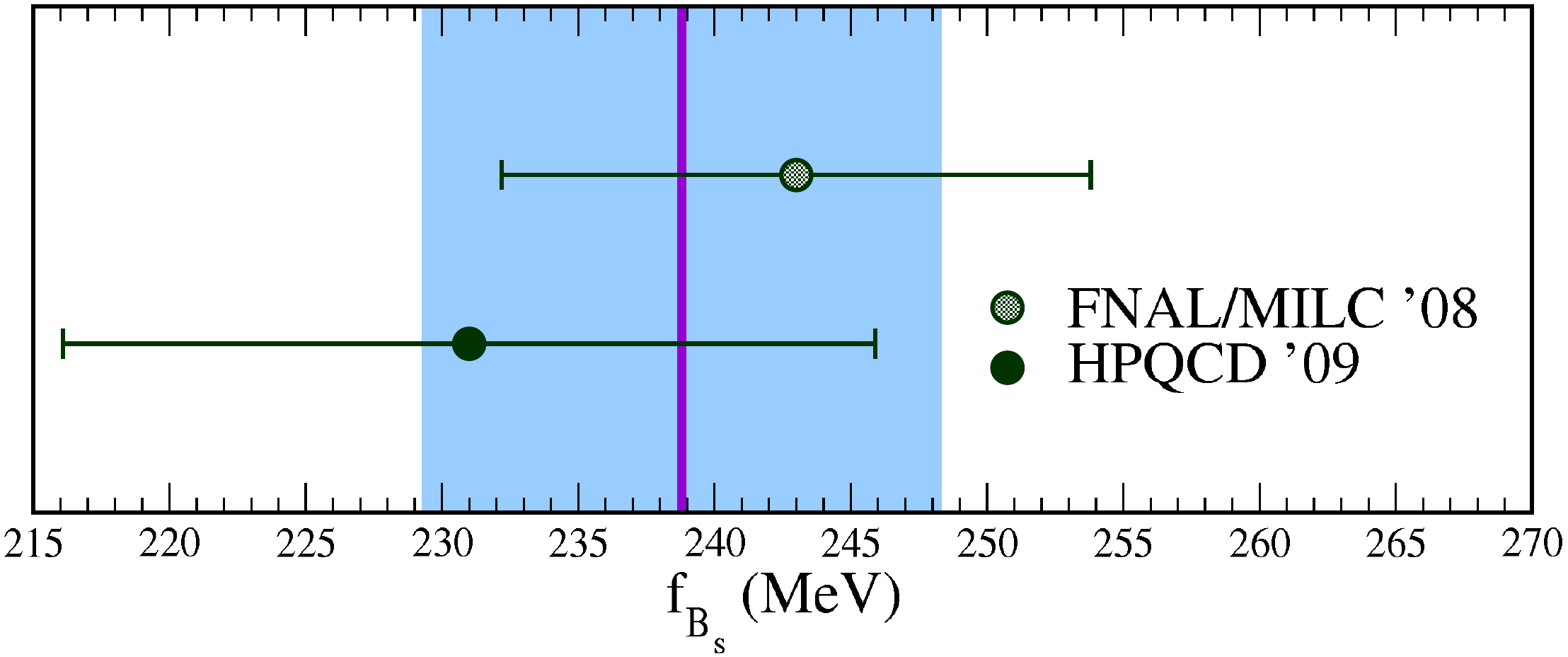}
\caption{Unquenched lattice average of the leptonic decay constant $f_{B_s}$.
The two $N_f = 2+1$ lattice inputs are given by shaded green circles with error bars, while the resulting average is denoted as a purple vertical line with a blue error band. 
\label{fig:fBs}}
\end{center}
\end{figure}

\begin{figure}[t]
\begin{center}
\includegraphics[width=0.6\linewidth]{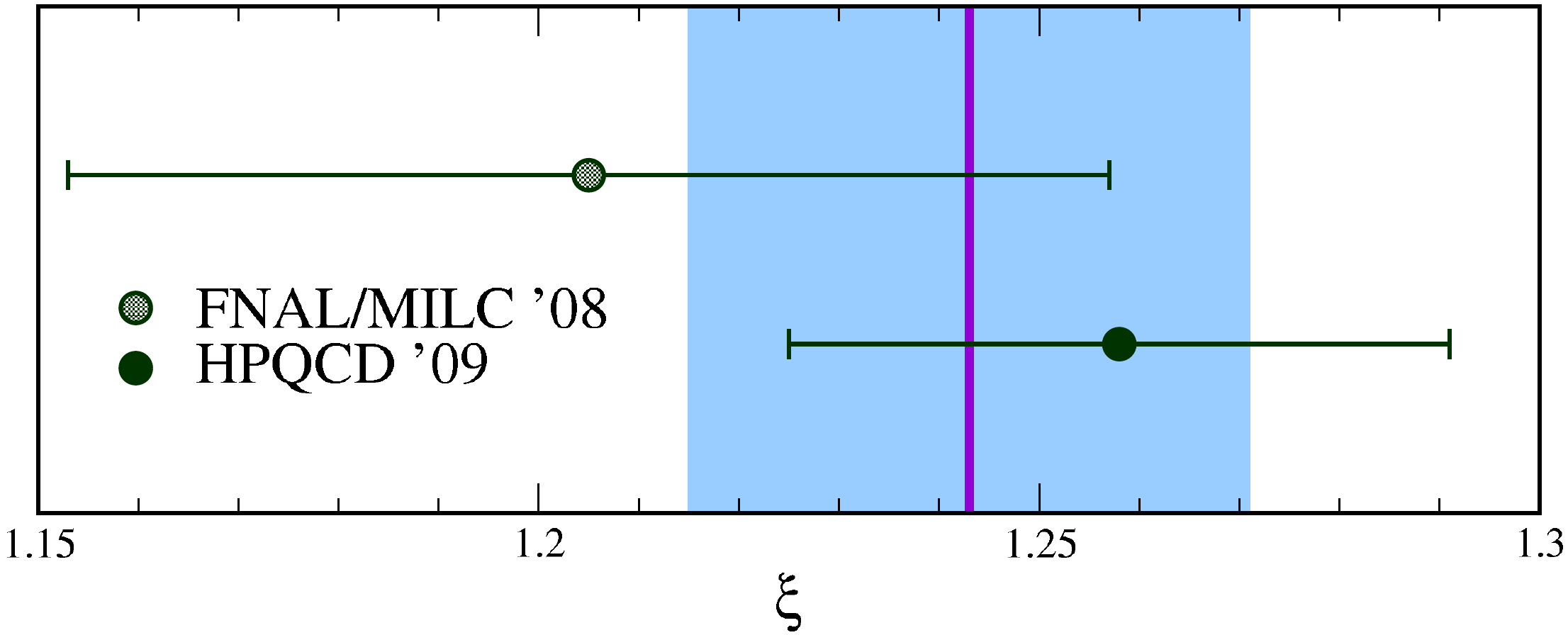}
\caption{Unquenched lattice average of the $SU(3)$-breaking ratio $\xi$.  The two $N_f = 2+1$ lattice inputs are given by shaded green circles with error bars, while the resulting average is denoted as a purple vertical line with a blue error band.
\label{fig:xi}}
\end{center}
\end{figure}

\begin{figure}[t]
\begin{center}
\includegraphics[width=0.6 \linewidth]{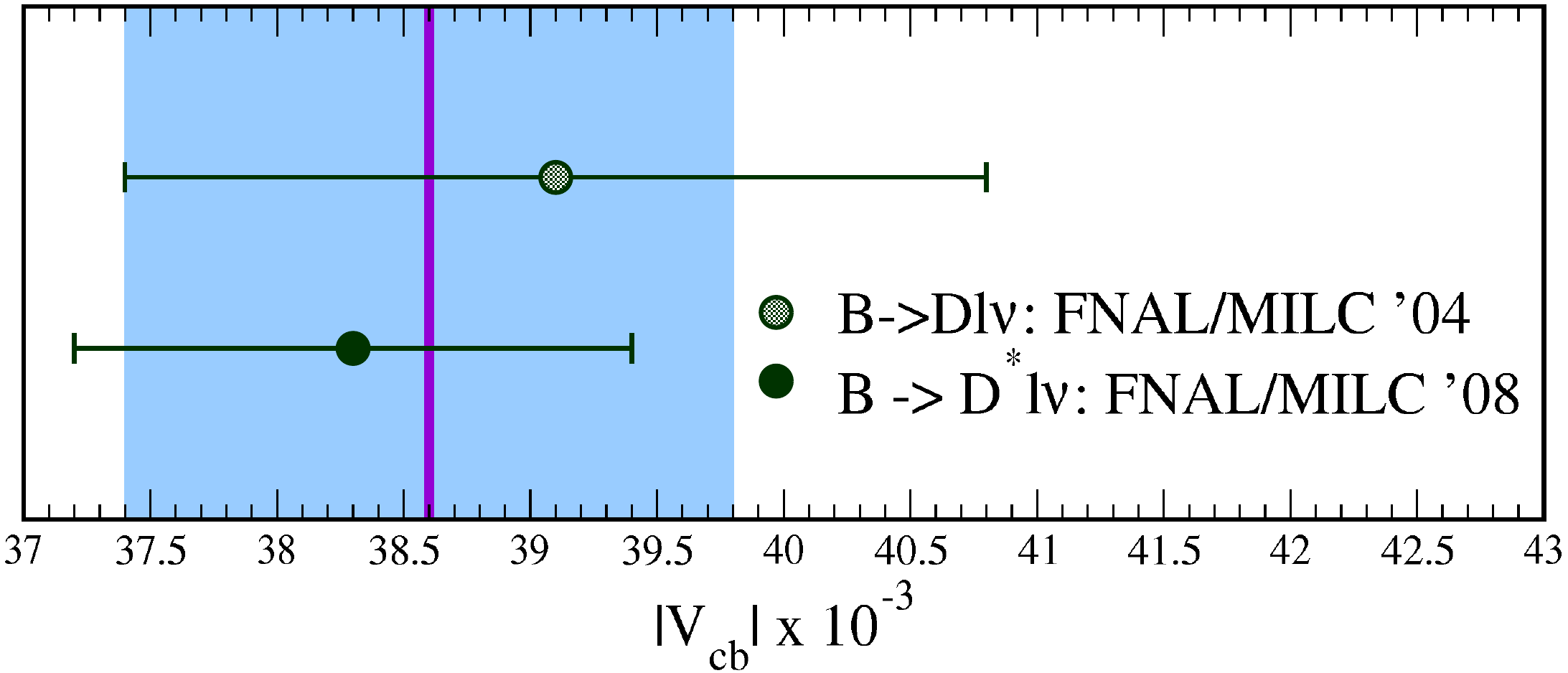}
\caption{Unquenched lattice average of the CKM matrix element $|V_{cb}|$.  The two $N_f = 2+1$ lattice inputs are given by shaded green circles with error bars, while the resulting average is denoted as a purple vertical line with a blue error band.
\label{fig:Vcb}}
\end{center}
\end{figure}

\begin{figure}[t]
\begin{center}
\includegraphics[width=0.6\linewidth]{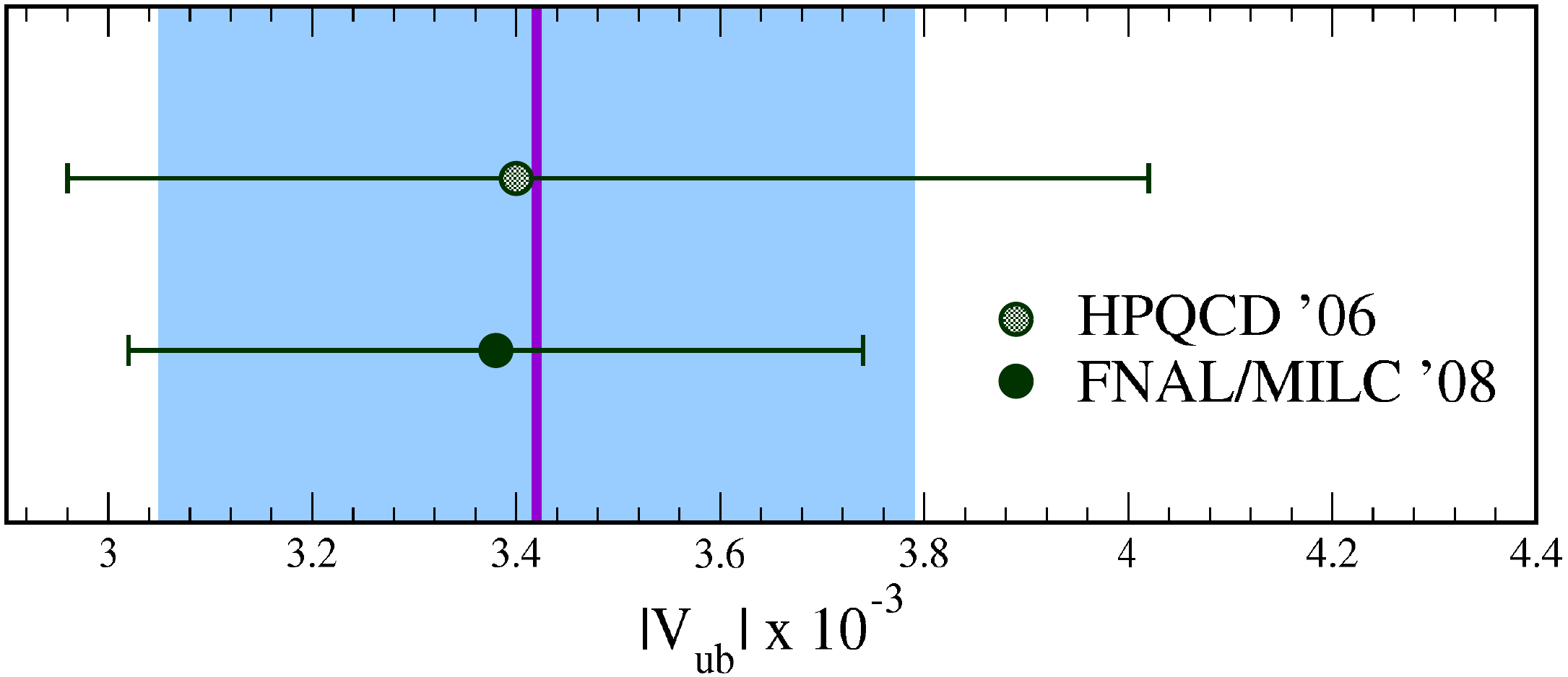}
\caption{Unquenched lattice average of the CKM matrix element $|V_{ub}|$.  The two $N_f = 2+1$ lattice inputs are given by shaded green circles with error bars, while the resulting average is denoted as a purple vertical line with a blue error band. 
\label{fig:Vub}}
\end{center}
\end{figure}

\begin{figure}[t]
\begin{center}
\includegraphics[width=0.6\linewidth]{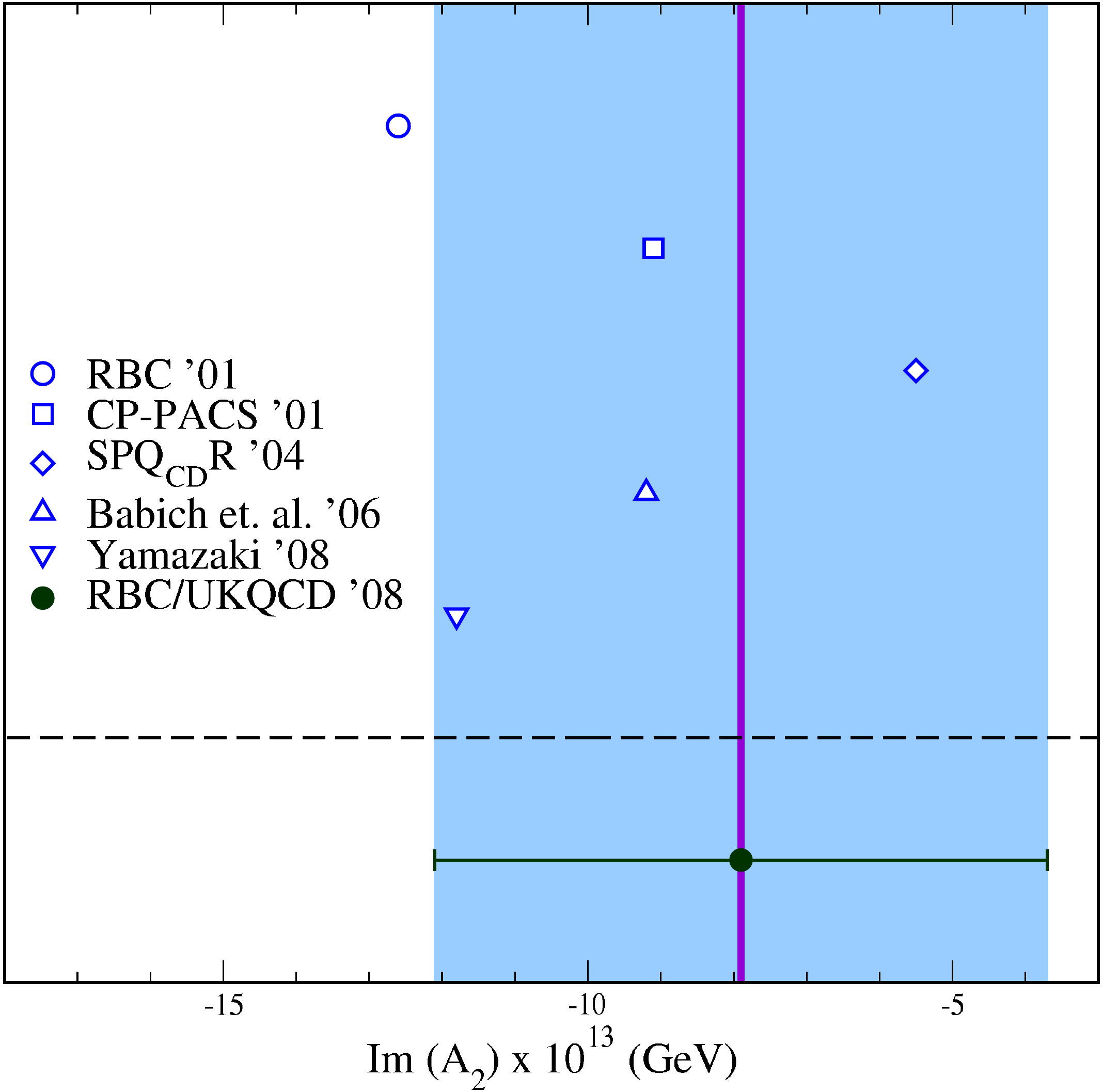}
\caption{Lattice determinations of the $K \to \pi \pi$ matrix element $\rm{Im}(A_2)$.  The only available $N_f = 2+1$ lattice determination is given by a filled green circles with error bars.  For comparison, the results of several quenched calculations (open blue symbols) are overlaid on the unquenched result (purple vertical line with blue error band).
\label{fig:ImA2}}
\end{center}
\end{figure}

\begin{figure}[t]
\begin{center}
\includegraphics[width=0.6\linewidth]{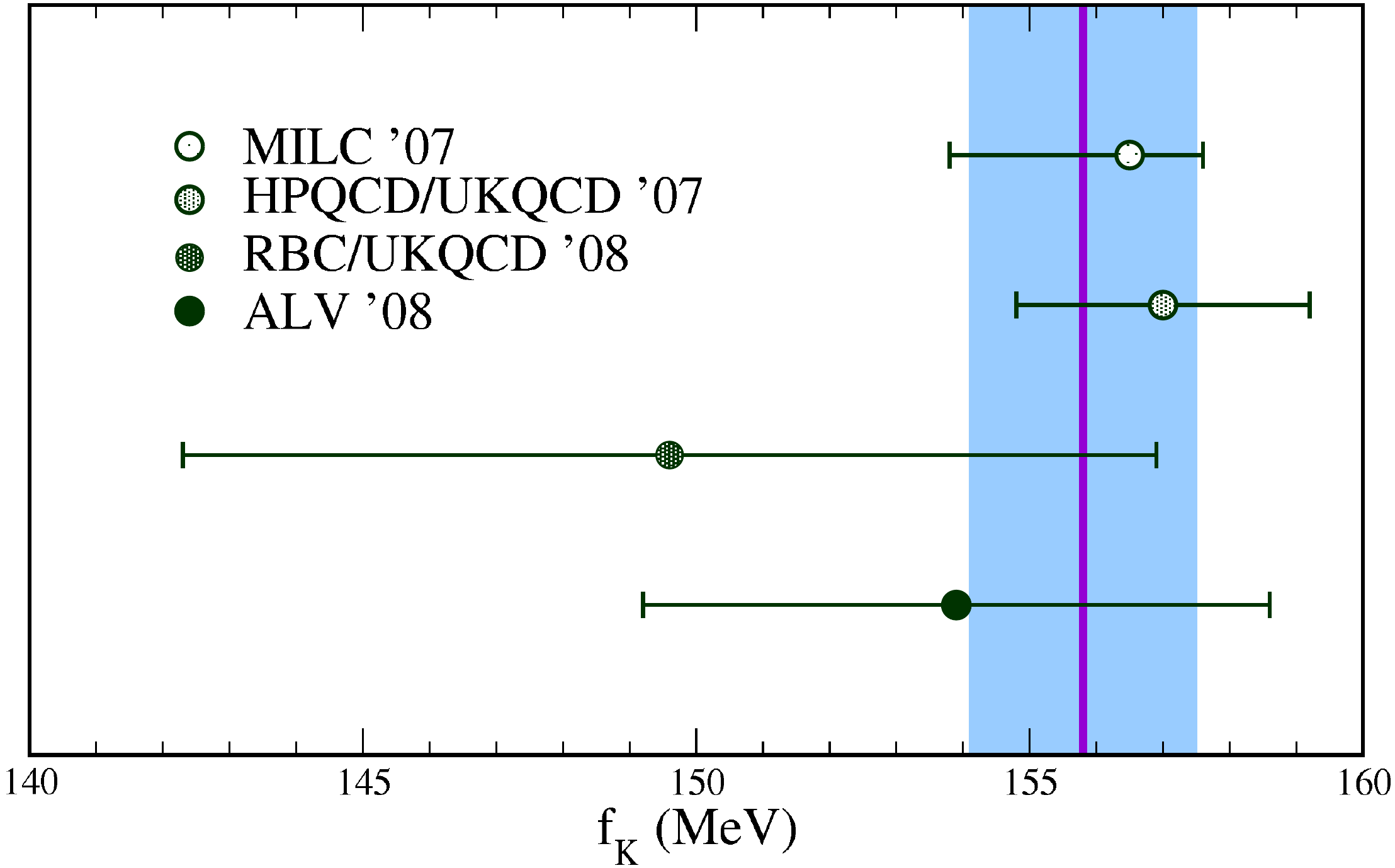}
\caption{Unquenched lattice average of the leptonic decay constant $f_K$.  The four $N_f = 2+1$ lattice inputs are given by shaded green circles with error bars, while the resulting average is denoted as a purple vertical line with a blue error band.  
\label{fig:fK}}
\end{center}
\end{figure}

%=================================================
%The bibliography
%=================================================

\bibliography{UTfit-hepv2}

\end{document}